\documentclass[twocolumn,english,aps,prb,floatfix]{revtex4-1}
\usepackage[T1]{fontenc}
\usepackage[latin9]{inputenc}
\usepackage{amsmath}
\usepackage{amssymb}
\usepackage{graphicx}
\usepackage{esint}

\def\s{{e^{i\pi\hat{S}}}}
\def\q{{e^{i\pi\hat{Q}}}}
\def\st{{e^{i\pi\hat{S}_3}}}
\def\qt{e^{i\pi\hat{Q}_3}}
\def\qo{e^{i\pi\hat{Q}_1}}
\def\qtw{e^{i\pi\hat{Q}_2}}

\def\so{e^{i\pi\hat{S}_1}}
\def\stw{e^{i\pi\hat{S}_2}}

\def\9{{\rangle}}
\def\6{{\langle}}

\def\beq{\begin{equation}}
\def\eeq{\end{equation}}
\def\bea{\begin{eqnarray}}
\def\eea{\end{eqnarray}}

\def\6{\langle}
\def\9{\rangle}
\makeatletter

\providecommand{\tabularnewline}{\\}

\@ifundefined{textcolor}{}
{%
 \definecolor{BLACK}{gray}{0}
 \definecolor{WHITE}{gray}{1}
 \definecolor{RED}{rgb}{1,0,0}
 \definecolor{GREEN}{rgb}{0,1,0}
 \definecolor{BLUE}{rgb}{0,0,1}
 \definecolor{CYAN}{cmyk}{1,0,0,0}
 \definecolor{MAGENTA}{cmyk}{0,1,0,0}
 \definecolor{YELLOW}{cmyk}{0,0,1,0}
 }

\usepackage{graphics}\usepackage{dcolumn}\usepackage{bm}

\makeatother

\usepackage{babel}

\begin{document}


\title{Fractionalizing Majorana fermions: non-abelian statistics on the edges of abelian quantum Hall states}


\author{Netanel H. Lindner}

\thanks{These authors have contributed equally to this work.}

\affiliation{Institute of Quantum Information and Matter, California
Institute of Technology, Pasadena, CA 91125, USA}

\affiliation{Department of Physics, California Institute of
Technology, Pasadena, CA 91125, USA}

\author{Erez Berg}

\thanks{These authors have contributed equally to this work.}

\affiliation{Department of Physics, Harvard University, Cambridge,
MA 02138, USA}

\author{Gil Refael}

\affiliation{Department of Physics, California Institute of
Technology, Pasadena, CA 91125, USA}

\author{Ady Stern}

\affiliation{Department of Condensed Matter Physics, Weizmann
Institute of Science, Rehovot 76100, Israel}

\date{\today}

\begin{abstract}
We study the non-abelian statistics characterizing systems where
counter-propagating gapless modes on the edges of fractional
quantum Hall states are gapped by proximity-coupling to
superconductors and ferromagnets. The most transparent example is
that of a fractional quantum spin Hall state, in which electrons
of one spin direction occupy a fractional quantum Hall state of
$\nu= 1/m$, while electrons of the opposite spin occupy a similar
state with $\nu = -1/m$. However, we also propose other examples
of such systems, which are easier to realize experimentally. We
find that each interface between a region on the edge coupled to a
superconductor and a region coupled to a ferromagnet corresponds
to a non-abelian anyon of quantum dimension $\sqrt{2m}$. We
calculate the unitary transformations that are associated with
braiding of these anyons, and show that they are able to realize a
richer set of non-abelian representations of the braid group than
the set realized by non-abelian anyons based on Majorana fermions.
We carry out this calculation both explicitly and by applying
general considerations. Finally, we show that topological
manipulations with these anyons cannot realize universal quantum
computation.

\end{abstract}

\maketitle

\section{Introduction}

\begin{figure}[b]
\centerline{\includegraphics[width=1\columnwidth]{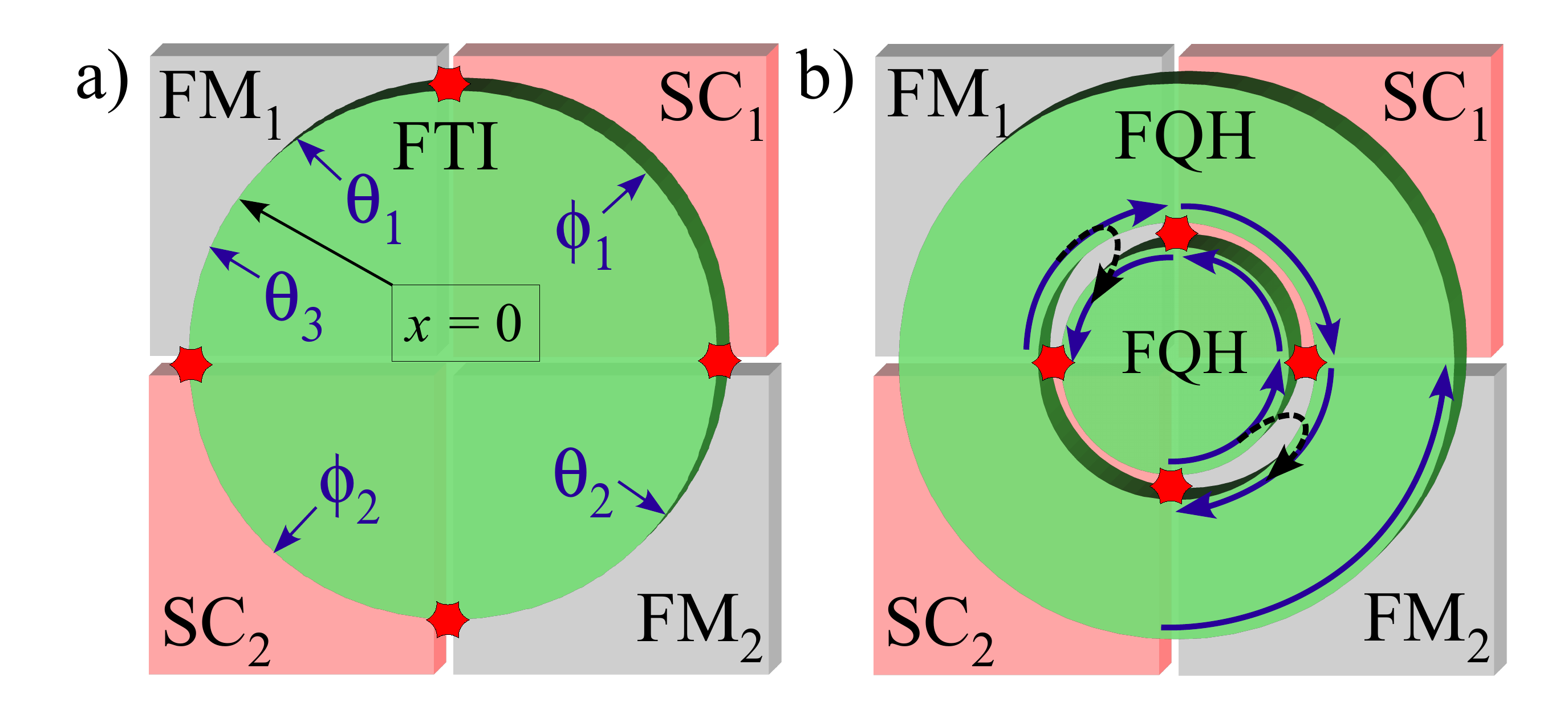}}\caption{\textbf{Schematic
setup.} (a) A fractional topological insulator (FTI) realization.
A FTI droplet with an odd filling factor $1/m$ is proximity
coupled to ferromagnets (FM) and to superconductors (SC), which
gap out its edge modes. The interfaces between the SC and FM
segments on the edge of the FTI are marked by red stars. (b) A
fractional quantum Hall (FQH) realization. A FQH droplet with
filling factor $1/m$ is separated by a thin barrier into two
pieces: an inner disk, and an outer annulus. On either side of the
barrier, there are counter-propagating edge states, which are
proximity coupled to superconductors and ferromagnets.}
\label{fig:system}
\end{figure}

Recent years have witnessed an extensive search for electronic
systems in which excitations (``quasi-particles'') follow
non-abelian quantum statistics. In such systems, the presence of
quasi-particles, also known as ``non-abelian anyons''
\cite{Lienass, Blok, SternInsight}, makes the ground state
degenerate. A mutual adiabatic interchange of quasi-particles'
positions\cite{Arovas} implements a unitary transformation that
operates within the subspace of ground states, and shifts the
system from one ground state to another. Remarkably, this unitary
transformation depends only on the topology of the interchange,
and is insensitive to imprecision and noise. These properties make
non-abelian anyons a test-ground for the idea of topological
quantum computation\cite{Kitaev2003}. The search for non-abelian
systems originated from the Moore-Read theory \cite{MooreRead} for
the $\nu=5/2$ Fractional Quantum Hall (FQH) state, and went on to
consider other quantum Hall states \cite{ReadRezayi,SternAnnals},
spin systems \cite{KitaevHex}, $p$-wave superconductors
\cite{ReadGreen,Ivanov, NayakWilczek}, topological insulators in
proximity coupling to superconductors \cite{FuPRL, FuPRB} and
hybrid systems of superconductors coupled to semiconductors where
spin-orbit coupling is strong
\cite{Sau2010,ZhangQHSC,SauSpinOrbit,Stanescu,OregWire,LutchynWire,
Marcel}. Signatures of Majorana zero modes may have been observed
in recent experiments \cite{Leopaper, Willett, Kang, Rokhinson,
Das}.

In the realizations based on superconductors, whether directly or
by proximity, the non-abelian statistics results from the
occurrence of zero-energy Majorana fermions bound to the cores of
vortices or to the ends of one dimensional wires
\cite{ReadGreen,Ivanov,NayakWilczek, FuPRL,FuPRB, Sau2010,ZhangQHSC, SauSpinOrbit,Stanescu,OregWire,LutchynWire, Marcel, JasonReview,BeenakkerReview}. Majorana-based
non-abelian statistics is, from the theory side, the most solid
prediction for the occurrence of non-abelian statistics, since it
is primarily based on the well tested BCS mean field theory of
superconductivity. Moreover, from the experimental side it is the
easiest realization to observe\cite{Leopaper}. The set of unitary
transformations that may be carried out on Majorana-based systems
is, however, rather limited, and does not allow for universal
topological quantum computation \cite{RMP,FreedmanLarsen}.

In this work we introduce and analyze a non-abelian system that is
based on proximity coupling to a superconductor but goes beyond
the Majorana fermion paradigm. The system we analyze is based on
the proximity-coupling of fractional quantum Hall systems or
fractional quantum spin Hall systems \cite{LevinFTI}  to
superconductors and ferromagnetic insulators (we will use the
terms ``fractional topological insulators'' and ``fractional
quantum spin Hall states'' interchangeably). The starting point of
our approach is the following observation,  made by Fu and Kane
\cite{FuPRB} when considering the edge states of two-dimensional
(2D) topological insulators of non-interacting electrons, of which
the integer quantum spin Hall state \cite{Kane, Bernevig} is a
particular example: In a 2D topological insulator, the gapless
edge modes may be gapped either by breaking time reversal symmetry
or by breaking charge conservation along the edge. The former may
be broken by proximity coupling to a ferromagnet, while the latter
may be broken by proximity coupling to a superconductor.
Remarkably, there must be a single Majorana mode localized at each
interface between a region where the edge modes are gapped by a
superconductor to a region where the edge modes are gapped by a
ferromagnet.

Our focus is on similar situations in cases where the gapless edge
modes are of fractional nature. We find that under these
circumstances, the Majorana operators carried by the interfaces in
the integer case are replaced by ``fractional Majorana operators''
whose properties we study.

We consider three types of physical systems. The first (shown
schematically in Fig. \ref{fig:system}a) is that of a 2D
fractional topological insulator\cite{LevinFTI}, that may be
viewed as a 2D system in which electrons of spin-up form an FQH
state of a Laughlin \cite{Laughlin} fraction $\nu=1/m$ , with $m$
being an odd integer, and electrons of spin-down form an FQH state
of a Laughlin fraction $\nu=-1/m$.

The second system (\ref{fig:system}b) is a Laughlin FQH droplet of
$\nu=1/m$, divided by a thin insulating barrier into an inner disk
and an outer annulus.
On the inner disk, the electronic spins are polarized parallel to
the
magnetic field (spin-up), and 
on the annulus the electronic spins are polarized anti-parallel to
the magnetic field (spin-down). 
Consequently, two edge modes flow on the two sides of the barrier,
with opposite spins and opposite velocities. Such a state may be
created under circumstances where the sign of the $g$-factor is
made to vary across the barrier. 

The third system is an electron-hole bi-layer subjected to a
perpendicular magnetic field, in which one layer is tuned to an
electron spin-polarized filling factor of $\nu=1/m$, and the other
to a
 hole spin-polarized $\nu=1/m$ state. In particular, this may be realized in a material with a spectrum that is electron-hole symmetric, such as graphene.

In all these cases, the gapless edge mode may be gapped either by
proximity coupling to a superconductor or by proximity coupling to
a ferromagnet. We imagine that the edge region is divided into
$2N$ segments, where the superconducting segments are all
proximity coupled to the same bulk superconductor, and the
ferromagnetic segments are all proximity coupled to the same
ferromegnet. The length of each segment is large compared to the
microscopic lengths, so that tunneling between neighboring SC-FM
interfaces is suppressed. We consider the proximity interactions
of the segments with the superconductor and the ferromegnet to be
strong.

The questions we ask ourselves are motivated by the analogy with
the non-interacting systems of Majorana fermions: what is the
degeneracy of the ground state? Is this degeneracy topologically
protected? What is the nature of the degenerate ground states? And
how can one manipulate the system such that it evolves, in a
protected way, between different ground states?

The structure of the paper is as follows: In
Sec.~\ref{sec:physical-picture} we give the physical picture that
we developed, and summarize our results. In
Sec.~\ref{sec:edge-model} we define the Hamiltonian of the system.
In Sec.~\ref{sec:Ground-state-degeneracy} we calculate the ground
state degeneracy. In Sec.~\ref{sec:domain-wall-operators} we
define the operators that are localized at the interfaces, and act
on the zero energy subspace. In
Sec.~\ref{sec:topological-manipulations} we calculate in detail
the unitary transformation that corresponds to a braid operation.
In Sec.~\ref{sec: TS} we show how this transformation may be
deduced from general considerations, bypassing the need for a
detailed calculation. In Sec.~\ref{sec:quantum-computing} we
discuss several aspects of the fractionalized Majorana operators,
and their suitability for topological quantum computation.
Sec.~\ref{sec:conclusions} contains some concluding remarks. The
paper is followed by appendices which discuss several technical
details.

\section{The physical picture and summary of the results}
\label{sec:physical-picture}

There are  three types of regions in the systems we consider: the
bulk, the parts of the edge that are proximity-coupled to a
superconductor, and the parts of the edge that are
proximity-coupled to a ferromagnet.

The bulk is either a fractional quantum Hall state or a fractional
quantum spin Hall state. In both cases it is gapped and
incompressible, and its elementary excitations are localized
quasi-particles whose charge is a multiple of  $e^*=e/m$ electron
charges. In our analysis we will assume that the area enclosed by
the edge modes encloses $n_\uparrow$ quasi-particles of spin-up
and $n_\downarrow$ quasi-particles of spin down. These
quasi-particles are assumed to be immobile.

In the parts of the edge that are coupled to a superconductor the
charge is defined only modulo $2e$, because Cooper-pairs may be
exchanged with the superconductor. Thus, the proper operator to
describe the charge on a region of this type is $e^{i\pi
\hat{Q}_i}$, with $\hat{Q}_i$ being the charge in $i$'th superconducting region.
Since the superconducting region may exchange $e^*$ charges with
the bulk, these operators may take the values $e^{i\pi q_i/m}$,
with $q_i$ an integer whose value is between zero and $2m-1$. The
pairing interaction leads to a ground state that is a spin
singlet, and thus the expectation value of the spin within each
superconducting region vanishes. As we show below, the Hamiltonian
of the system commutes with the operators $e^{i\pi \hat{Q}_i}$ in
the limit we consider. For the familiar $m=1$ case, these
operators measure the parity of the number of electrons within
each superconducting region.

The edge regions that are proximity-coupled to ferromagnets are,
in some sense, the dual of the superconducting regions. The
ferromagnet introduces back-scattering between the two
counter-propagating edge modes, leading to the formation of an
energy gap. If the chemical potential lies within this gap, the
region becomes insulating and incompressible. Consequently, the
charge in the region does not fluctuate, and its value may be
defined as zero. The spin, on the other hand, does fluctuate.
Since the back-scattering from spin up electron to spin down
electron changes the total spin of the region by two (where the
electronic spin is defined as one unit of spin), the operator that
may be expected to have an expectation value within the ground
state is $e^{i\pi \hat{S}_i}$, with $\hat{S}_i$ being the total
spin in the $i$'th ferromagnet region. Again, spins of $1/m$ may
be exchanged with the bulk, and thus these operators may take the
eigenvalues $e^{i\pi s_i/m}$, with $s$ an integer between zero and
$2m-1$. The Hamiltonian of the system commutes with the operators
$e^{i\pi
\hat{S}_i}$ in the limit we consider.

The operators $e^{i\pi \hat{Q}_i}$ and $e^{i\pi \hat{S}_i}$ label
the different domains in the system, as indicated in
Fig~\ref{fig:braiding}a. They satisfy a constraint dictated by the
state of the bulk,
\begin{eqnarray}
\prod_{i=1}^{N} e^{i\pi \hat{Q}_i}=e^{i\pi(n_\uparrow+n_\downarrow)/m}\nonumber \\
\prod_{i=1}^{N} e^{i\pi
\hat{S}_i}=e^{i\pi(n_\uparrow-n_\downarrow)/m} \label{bcqs}
\end{eqnarray}
For the familiar $m=1$ case there are only two possible solutions
for these constraints, corresponding to the two right hand sides
of Eq. (\ref{bcqs}) being both $+1$ or both $-1$. For a general
$m$, the number of topologically distinct constraints is  $2m^2$,
since Eqs. (\ref{bcqs}) are invariant under the transformation
where $n_\uparrow\rightarrow n_\uparrow \pm m$ together with
$n_\downarrow\rightarrow n_\downarrow \pm m$. These sets may be
spanned by the values $0\le n_\uparrow\le 2m-1$ and $0\le
n_\downarrow \le m-1$.

The degeneracy of the ground state may be understood by examining
the algebra constructed by the operators $e^{i\pi \hat{Q}_i}$ and
$e^{i\pi \hat{S}_i}$.
As we show in the next section, the operators $e^{i\pi \hat{S}_i},
e^{i\pi{ \hat{Q}}_i}$ satisfy
\begin{eqnarray}
\nonumber [e^{i\pi { \hat{Q}}_i},e^{i\pi { \hat{Q}}_j}]&=&[e^{i\pi { \hat{S}}_i},e^{i\pi { \hat{S}}_j}]=0,\\
\nonumber \left[ e^{i\pi {\hat{Q}_j}}, \prod_{i=1}^N e^{i\pi \hat{S}_i}\right] &=& \left[e^{i\pi {\hat{S}_j}}, \prod_{i=1}^N e^{i\pi \hat{Q}_i}\right] = 0, \\
e^{i\pi \hat{Q}_j}e^{i\pi {\sum_{k=1}^l
\hat{S}}_k}&=&e^{i\frac{\pi}{m}\delta_{jl}} e^{i\pi {\sum_{k=1}^l
\hat{S}}_k}e^{i\pi \hat{Q}_j}, \label{algebra}
\end{eqnarray}
where, in the last equation, $1\le j,l<N$ (see
Fig~\ref{fig:braiding}a for the enumeration convention). As
manifested by Eqs. (\ref{algebra}), the pairs of operators
$e^{i\pi \hat{Q}_i}, e^{i\pi {\sum_{k=1}^i \hat{S}}_k}$ form $N-1$
pairs of degrees of freedom, where members of different pairs
commute with one another. 
It is the relation between members of the same pairs, expressed in
Eq. (\ref{algebra}), from which the ground state degeneracy may be
easily read out. As is evident from this equation, if
$|\psi\rangle$ is a ground state of the system which is also an
eigenstate of $e^{i\pi \hat{Q}_j}$, then $2m-1$ additional ground
are $\left (e^{i\pi \sum_{i=1}^{j}\hat{S}_i}\right
)^k|\psi\rangle$, where $k$ is an integer between $1$ and $2m-1$.
With $N-1$ mutually independent pairs, we reach the conclusion
that the ground state degeneracy, for a given value of
$n_\uparrow, n_\downarrow$, is $(2m)^{N-1}$.

The operators acting within a sector of given $n_\uparrow$,
$n_\downarrow$ of the ground state subspace are represented by
$(2m)^{N-1}\times (2m)^{N-1}$ matrices. They may be expressed in
terms of sums
and products of the operators appearing in (\ref{algebra}). 
The physical operations described by the operators $e^{i\pi
\hat{S}_i}, e^{i\pi{\hat{Q}}_i}$ can also be read off the
relations (\ref{algebra}). The operator $e^{i\pi \hat{S}_i}$
transfers a quasi-particle of charge $e/m$ from the $i-1$'th
superconductor to the $i$'th superconductor. Since the spin within
the superconductor vanishes, there is no distinction, within the
ground state manifold, between the possible spin states of the
transferred quasi-particle. In contrast, the operator
$e^{i\pi\sigma \hat{Q}_i}$ transfers a quasi-particle of
spin $\sigma=\pm 1$ across the $i$'th superconductor. 

For $m=1$ the operators $e^{i\pi \hat{S}_i}$ and $e^{i\pi
\hat{Q}_i}$, measuring the parity of the spin and the charge in
the $i$'th ferromagnetic and superconducting region, respectively,
may be expressed in terms of Majorana operators that reside at the
interfaces bordering that region. A similar representation exists
also in the case of $m\ne 1$. Its details are given in Section
\ref{sec:domain-wall-operators}.


We stress that the ground state degeneracy is \emph{topological},
in the sense that no measurement of a local operator can determine
the state of the system within the ground state subspace. For
$m=1$, this corresponds to the well-known ``topological
protection'' of the ground state subspace of Majorana fermions
\cite{RMP,SternFelix}, as long as single electron tunneling is
forbidden either between the different Majorana modes, or between
the Majorana modes and the external world. In the fractional case
the states in the ground state manifold can be labelled by the
fractional part of the spin or charge of the FM/SC segments,
respectively. These clearly cannot be measured locally. Moreover,
they can only change by tunneling fractional quasi-particles
between different segments; even tunneling electrons from the
outside environment cannot split the degeneracy completely,
because it can only change the charge and spin of the system by
integers.

Topological manipulations of non-abelian anyons confined to one
dimension are somewhat more complicated than those carried out in
two dimensions. The simplest manipulation does not involve any
motion of the anyons, but rather involves either a $2\pi$ twist of
the order parameter of the superconductor coupled to one or
several superconducting segments, or a $2\pi$ rotation of the
direction of the magnetization of the ferromagnet coupled to the
insulating segments\cite{TeoKane}. When  a vortex encircles the
$i$'th superconducting region it leads to the accumulation of a
Berry phase of $2\pi$ multiplied by the number of Cooper-pairs it
encircles. In the problem we consider, this phase amounts to $e^{i
\pi \hat{Q}_i}$, and that is the unitary transformation applied by
such rotation. As explained above, this transformation transfers a
spin of $1/m$ between the two ferromagnetic regions with which the
superconductor borders. Similarly, a rotation of the magnetization
in the ferromagnetic region leads to a transfer of a charge of
$e/m$ between the two superconductors with which the ferromagnet
borders.

A more complicated manipulation is that of anyons' braiding, and
its associated non-abelian statistics. While in two dimensions the
braiding of anyons is defined in terms of world lines ${\bf R}(t)$
that braid one another as time evolves, in one dimension - both in
the integer $m=1$ and in the fractional case - a braiding
operation requires the introduction of tunneling terms between
different points along the edge\cite{AliceaBraiding, Sau2011}. The
braiding is then defined in terms of trajectories in parameter
space, which
includes the tunneling amplitudes that are introduced to implement the braiding. 
The braiding is topological in the sense that it does not depend
on the precise details of the trajectory that implements it, as
long as the degeneracy of the ground state manifold does not vary
throughout the implementation.
Physically, one can imagine realizing such operations by changing
external gate potentials which deform the shape of the system's
edge adiabatically (similar to the operations proposed for the
Majorana case \cite{AliceaBraiding, Alicea2012}).

In the integer $m=1$ case, the interchange of two anyons
positioned at two neighboring interfaces is carried out by
subjecting the system to an adiabatically time dependent
Hamiltonian in which interfaces are coupled to one another. When
two or three interfaces are coupled to one another, the degeneracy
of the ground state does not depend on the precise value of the
couplings, as long as they do not all vanish at once.
Consequently, one may ``copy'' an anyon $a$ onto an anyon $c$ by
starting with a situation where corresponding interfaces $b$ and
$c$ are tunnel coupled, and then turning on a coupling between $a$
and $b$ while simultaneously turning off the coupling of $b$ to
$c$. Three consecutive ``copying'' processes then lead to an
interchange,
and the resulting interchanges  generate a non-abelian
representation of the braid group.

In the integer case only electrons may tunnel between two
interfaces, thus allowing us to characterize the tunneling term by
one tunneling amplitude. In contrast, in the fractional case more
types of tunneling processes are possible, corresponding to the
tunneling of any number of quasi-particles of charges $e/m$ and
spin $\pm 1/m $. To define the effective Hamiltonian coupling two
interfaces, we need to specify the amplitudes for all these
distinct processes.
As one may expect, if only electrons are allowed to
tunnel between the interfaces (as may be the case if the tunneling
is constrained to take place through the vacuum), the $m=1$ case
is reproduced. When single quasi-particles of one spin direction
are allowed to tunnel (which is the natural case for the FQHE
realization of our model), tunnel coupling between either two or
three interfaces reduces the degeneracy of the ground state by a
factor of $2m$. This case then opens the way for interchanges of
the positions of anyons by the same method envisioned for the
integer case. We analyze these interchanges in detail below.


Our analysis of the unitary transformations that correspond to
braiding schemes goes follows different routes. In the first,
detailed in Section \ref{sec:topological-manipulations}, we
explicitly calculate these transformation for a particular case of
anyons interchange. In the second, detailed in Section \ref{sec:
TS}, we utilize general properties of anyons to find all
non-abelian representations of the braid group that satisfy
conditions that we impose, which are natural to expect from the
system we analyze. Both routes indeed converge to the same result.
While the details of the calculations are given in the following
sections, here we discuss their results.

To consider braiding, we imagine that two anyons at the two ends
of the $i$'th superconducting region are interchanged. For the
$m=1$ case the interchange of two Majorana fermions correspond to
the transformation
\begin{equation}
\frac{1}{\sqrt{2}}\left[1\pm i \exp(i\pi \hat{Q}_i)\right].
\label{majexchange}
\end{equation}
 This
transformation may be written as $\exp[i\frac{\pi}{2}
(\hat{Q}_i-k)^2]$ with $k=0,1$ corresponding to the $\pm$ sign in
(\ref{majexchange}), or as $\frac{1}{\sqrt{2}}\left
(1\pm\gamma_1\gamma_2\right )$, with $\gamma_1,\gamma_2$ the two
localized Majorana modes at the two ends of the superconducting
region. Its square is the parity of the charge in the
superconducting region, and its fourth power is unity. Note that
in two dimensions, the two signs in (\ref{majexchange}) correspond
to anyons exchange in clockwise and anti-clockwise sense. In
contrast, in one dimension the two signs may be realized by
different choices of tunneling amplitudes, and are not necessarily
associated with a geometric notion. Consistent with the
topological nature of the transformation, a trajectory that leads
to one sign in (\ref{majexchange}) cannot be deformed into a
trajectory that corresponds to a different sign, without passing
through a trajectory in which the degeneracy of the ground state
varies during the execution of the braiding.

Guided by this familiar example, we expect that at the fractional
case the unitary transformation corresponding to this interchange
will depend only on $e^{i\pi \hat{Q}_i}$. We expect to be able to
write it as
\begin{equation}
U(\hat{Q}_i)=\sum_{j=0}^{2m-1}a_j\exp{\left (i\pi j
\hat{Q}_i\right) }, \label{unitrans}
\end{equation}
with some complex coefficients $a_j$, i.e., to be periodic in
$\hat{Q}_i$, with the period being $2$. We expect the values of
$a_j$ to depend on the type of tunneling amplitudes that are used
to implement the braiding.

In our analysis,  we find a more compact, yet equivalent, form for
the transformation $U$, which is
\begin{equation}
U(\hat{Q}_i)=e^{i\alpha\pi \left( \hat{Q}_i-\frac{k}{m} \right
)^2}.
\label{unitranselegant}
\end{equation}
The value of $\alpha$ depends on the type of particle which
tunnels during the implementation of the braiding, while the value
of $k$ depends on the value of the tunneling amplitudes. For an
electron tunneling, $\alpha=\frac{m^2}{2}$. Just as for the $m=1$
case, for this value of $\alpha$ the unitary transformation
(\ref{unitranselegant}) has two possible eigenvalues,  $U^4=1$,
and it is periodic in $k$ with a period of $2$.  For braiding
carried out by tunneling single quasi-particles we find
$\alpha=\frac{m}{2}$. In this case $U^{4m}=1$, and $U$ is periodic
in $k$ with a period of $2m$. 

Just as in the $m=1$ case, trajectories in parameter space that
differ by their value of $k$ are separated by trajectories that
involve a variation in the degeneracy of the ground state. We note
that up to an unimportant abelian phase, the unitary
transformation (\ref{unitranselegant}) may be thought of as
composed of a transformation $e^{i\alpha\pi \hat{Q}_i^2}$ that
results from an interchange of anyons, multiplied by a
transformation $e^{\frac{2\alpha\pi i}{m} \hat{Q}_ik}$ that
results from a vortex encircling the $i$'th superconducting region
$2\alpha k/m$ times.

Non-abelian statistics is the cornerstone of topological quantum
computation \cite{RMP,Kitaev2003}, due to possibility it opens for
the implementation of unitary transformations that are
topologically protected from decoherence and noise. It is then
natural to examine whether the non-abelian anyons that we study
allow for universal quantum computation, that is, whether any
unitary transformation within the ground state subspace may be
approximated by topological manipulations of the anyons
\cite{FreedmanLarsen}. We find that, at least for unitary time
evolution (i.e., processes that do not involve measurements) the
answer to this question is negative, as it is for the integer
case.

\section{Edge model}\label{sec:edge-model}

The edge states of a FTI are described by a hydrodynamic bosonized
theory \cite{WenChiralLL,LeeEdges}. The edge effective Hamiltonian
is written as

\begin{align}
H & =\frac{m u}{2\pi}\int dx\left[K\left(x\right)\left(\partial_{x}\phi\right)^{2}+\frac{1}{K\left(x\right)}\left(\partial_{x}\theta\right)^{2}\right]\nonumber \\
 & -\int dx\left[g_{S}\left(x\right)\cos\left(2m\phi\right)+g_{F}\left(x\right)\cos\left(2m\theta\right)\right].\label{eq:H}
\end{align}
 Here, $u$ is the edge mode velocity, $\phi$, $\theta$ are bosonic
fields satisfying the commutation relation
$\left[\phi\left(x\right),\theta\left(x'\right)\right]=\frac{i\pi}{m}\Theta\left(x'-x\right)$
where $\Theta$ is the Heaviside step function, and
$g_{S}\left(x\right)$, $g_{F}\left(x\right)$ describe
position-dependent proximity couplings to a SC and a FM, which we
take to be constant in the SC/FM regions and zero elsewhere,
respectively. The magnetization of the FM is taken to be in the
$x$ direction. $K\left(x\right)$ is a space-dependent Luttinger
parameter, originating from interactions between electrons of
opposite spins. The charge and spin densities are given by
$\rho=\partial_{x}\theta/\pi$ and $s^{z}=\partial_{x}\phi/\pi$,
respectively (where the spin is measured in units of the electron
spin $\hbar/2$). A right or left moving electron is described by
the operators $\psi_{\pm}=e^{im\left(\phi\pm\theta\right)}$.

Crucially for the arguments below, we will assume that \emph{the
entire edge is gapped} by the proximity to the SC and FM, except
(possibly) the SC/FM interface. This can be achieved, in
principle, by making the proximity coupling to the SC and FM
sufficiently strong.

\section{Ground state degeneracy of disk with 2N segments\label{sec:Ground-state-degeneracy}}

We consider a disk with $2N$ FM/SC interfaces on its boundary
(illustrated in Fig. \ref{fig:system}a for $N=2$). In order to
determine the dimension of the ground state manifold, we construct
a set of commuting operators, which can be used to characterize
the ground states. Consider the operators: $e^{i\pi Q_{j}}\equiv
e^{i\left(\theta_{j+1}-\theta_{j}\right)}$, $j=1,\dots,N$, where
$\theta_{j}$ is a $\theta$ field evaluated at an arbitrary point
near the middle of the $j$th FM region. The origin ($x=0$) is
chosen to lie within the first FM region (see Fig.
\ref{fig:system}a). The operator $\theta_{N+1}$ is located within
this region, to the left of the origin ($x<0$), while $\theta_1$
is to the right of the origin ($x>0$). The fields $\theta$, $\phi$
satisfy the boundary conditions $e^{i\pi Q_{\mathrm{tot}}}=
e^{i\left[\theta(L^{-})-\theta(0^{+})\right]}$ and $e^{i\pi
S_{\mathrm{tot}}}= e^{i\left[\phi(L^{-})-\phi(0^{+})\right]}$,
where $L$ is the perimeter of the system, and $Q_{\mathrm{tot}}$,
$S_{\mathrm{tot}}$ are the total charge and spin on the edge,
respectively.

Since we are in the gapped phase of the sine-Gordon model of Eq.
(\ref{eq:H}), we expect in the thermodynamic limit (where the size
of \emph{all} of the segments becomes large) that the $\theta$
field is essentially pinned to the minima of the cosine potential
in the FM regions. (Similar considerations hold for the $\phi$
fields in the SC regions.) In other words, the
$\theta\rightarrow\theta+\pi/m$ symmetry is spontaneously broken.
In this phase, correlations of the fluctuations of $\theta$ decay
exponentially on length scales larger than the correlation length
$\xi\sim u/\Delta_{F}$, where $\Delta_{F}$ is the gap in the FM
regions (see Appendix~\ref{app:MatrixElements} for an analysis of
the gapped phase).  Therefore, one can construct
\emph{approximate} ground states which are characterized by
$\langle e^{i(\theta_{j+1}-\theta_j)} \rangle = \langle e^{i\pi
Q_{j}}\rangle\equiv\lambda_{j}\ne0$, where
$\lambda_{j}=\left|\lambda\right|e^{\frac{i\pi}{m}q_{j}}$, where
$q_{j}\in\left\{ 0,\dots,2m-1\right\}$ can be chosen independently
for each FM domain. The energy splitting between these ground
states is suppressed in the thermodynamic limit as $e^{-R/\xi}$,
where $R$ is the length each region, as discussed below and in
Appendix \ref{app:MatrixElements}.

In addition, $e^{i\pi S_{\mathrm{tot}}}$ commutes both with the
Hamiltonain and with $e^{i\pi Q_{j}}$. Therefore the ground states
can be chosen to be eigenstates of $e^{i\pi S_{\mathrm{tot}}}$,
with eigenvalues $e^{i\frac{\pi}{m}s}$, $s\in\left\{
0,\dots,2m-1\right\} $. We label the approximate ground states as
$\vert\left\{ q\right\} ;s\rangle\equiv\vert
q_{1},\dots,q_{N};s\rangle$, where $\vert\left\{ q\right\}
;s\rangle$ satisfies that $\langle\left\{ q\right\} ;s\vert
e^{i\pi Q_{j}}\vert\left\{ q\right\}
;s\rangle=\left|\lambda\right|e^{\frac{i\pi}{m}q_{j}}$.


For a large but finite system, the $\vert \{q\}; s \rangle$ states
are not exactly degenerate. There are two effects that lift the
degeneracy between them: intra-segment instanton tunneling events
between states with different $\{q\}$, and inter-segment
``Josephson'' couplings which make the energy dependent on the
values of $\{q\}$. However, both of these effects are suppressed
exponentially as $e^{-R/\xi}$, as they are associated with an
action which grows linearly with the system size. Therefore, we
argue that $\vert
\{q\}; s \rangle$ are approximately degenerate, up to
exponentially small corrections, for any choice of the set
$\{q\},s$.

Similarly, one can define a set of ``dual'' operators $e^{i\pi
S_{j}}\equiv e^{i\left(\phi_{j}-\phi_{j-1}\right)}$,
$j=2,\dots,N$, and $e^{i\pi S_{1}}=e^{i\pi S_{\mathrm{tot}}}
\prod^{N}_{i=2}e^{-i\pi S_i}$. Although the SC regions are in the
gapped phase, and the fields $\phi_{j}$ are pinned near the minima
of the corresponding cosine potentials, note that the approximate
ground states $\vert\left\{ q\right\} ;s\rangle$ \emph{cannot} be
further distinguished by the expectation values of the operators
$e^{i\pi S_{j}}$. In fact, these states satisfy
$\langle\{q\};s\vert e^{i\pi
S_{j}}\vert\{q\};s\rangle\rightarrow0$ in the thermodynamic limit.
That is because the operators $e^{i\pi S_{j}}$ and $e^{i\pi
Q_{j}}$ satisfy the commutation relations

\begin{equation}
e^{i\pi S_{i}}e^{i\pi
Q_{j}}=e^{\frac{i\pi}{m}\left(\delta_{i,j+1}-\delta_{i,j}\right)}e^{i\pi
Q_{j}}e^{i\pi S_{i}},\label{eq:comm}
\end{equation}
 which can be verified by using the commutation relation of the $\phi$
and $\theta$ fields. In the state $\vert\{q\};s\rangle$, the value
of $e^{i\pi Q_j}$ is approximately localized near
$e^{i\frac{\pi}{m}q_{j}}$. Applying the operator $e^{i\pi S_{j}}$
to this state shifts $e^{i\pi Q_j}$ to $e^{i\pi
(Q_j+\frac{1}{m})}$, as can be seen from Eq.~ (\ref{eq:comm}).
This shift implies that the overlap of the states
$\vert\{q\};s\rangle$ and $e^{i\pi S_{j}}\vert\{q\};s\rangle$
\emph{decays exponentially} with the system size.

Overall, there are $\left(2m\right)^{N+1}$ distinct approximate
eigenstate $\vert\{q\};s\rangle$, corresponding to the $2m$
allowed values of charges $q_{j}$ of each individual SC segment,
and the total spin $s$, which can also take $2m$ values. Not all
of these states, however, are physical.
Labelling the total charge by an integer $q =\sum_{j=1}^{N} q_j$,
we see from Eq.~(\ref{bcqs}) that $s$ and $q$ must be either both
even or both odd, corresponding to a total even or odd number of
fractional quasi-particles in the bulk of the system.
Due to this constraint, the number of \emph{physical} states is
only $\frac{1}{2} (2m)^{N+1}$.

In a given sector with a fixed total charge and total spin, there
are $N_{\mathrm{gs}}=\left(2m\right)^{N-1}$ ground states. For
$m=1$, we get $N_{\mathrm{gs}}=2^{N-1}$ for each parity sector, as
expected for $2N$ Majorana states located at each of the FM/SC
interfaces \cite{ReadGreen}.

The ground state degeneracy in the fractional case suggests that
each interface can be thought of as an anyon whose quantum
dimension is $\sqrt{2m}$. This is reminiscent of recently proposed
models in which dislocations in abelian topological phases carry
anyons with quantum dimensions which are square roots of
integers\cite{Bombin2010,Barkeshli2011,You2012}.

\section{Interface operators} \label{sec:domain-wall-operators}

We now turn to define physical operators that act on the
low-energy subspace. These operators are analogous to the Majorana
operators in the $m=1$ case, in the sense that they can be used to
express any physical observable in the low-energy subspace. They
will be useful when we discuss topological manipulations of the
low-energy subspace in the next section.

We define the unitary operators $e^{i\hat{\phi}_{i}}$ and
$e^{i\pi\hat{Q}_{j}}$ such that

\begin{equation}
e^{i\pi\hat{Q}_{j}}\vert q_{1},\dots,q_{N},s\rangle=e^{\frac{i\pi
q_{j}}{m}}\vert q_{1},\dots,q_{N},s\rangle,\label{eq:expQj}
\end{equation}

\begin{equation}
e^{i\hat{\phi}_{j}}\vert q_{1},\dots,q_{N},s\rangle=\vert
q_{1},\dots,q_{j}+1,\dots,q_{N},s\rangle.\label{eq:expPhij}
\end{equation}
 $e^{i\pi\hat{Q}_{j}}$ is a diagonal operator in the $\vert\{q\},s\rangle$
basis, whereas $e^{i\hat{\phi}_{j}}$ shifts $q_{j}$ by one. These
operators can be thought of as projections of the ``microscopic''
operators $e^{i\phi_{j}}$ and $e^{i\pi Q_{j}}$, introduced in the
previous section, onto the low-energy subspace. In addition, we
define the operator $\hat{T}_{s}$ that shifts the total spin of
the system:
\begin{equation}
\hat{T}_{s}\vert q_{1},\dots,q_{N},s\rangle=\vert q_{1},\dots,q_{N},s+1\rangle.\label{eq:Ts}
\end{equation}

The operators (\ref{eq:expPhij},\ref{eq:Ts}) will not be useful to
us, since they cannot be constructed
by projecting any 
combination of edge quasi-particle operators onto the low energy
subspace. To see this, note that they add a charge of $1/m$ and
zero spin or spin $1/m$ with no charge. As a result, they violate
the constraint between the total spin and charge,
Eq.~(\ref{bcqs}). However, these operators can be used to
construct the combinations
\begin{align}
\chi_{2j,\sigma} & =e^{i\hat{\phi}_{j}}(\hat{T}_{s})^{\sigma}\prod_{i=1}^{j}e^{i\sigma\pi\hat{Q}_{i}},\nonumber \\
\chi_{2j+1,\sigma} & =e^{i\hat{\phi}_{j+1}}(\hat{T}_{s})^{\sigma}
\prod_{i=1}^{j}e^{i\sigma\pi\hat{Q}_{i}},\label{eq:chi-def}
\end{align}
where $\sigma=\pm1$. These combinations, which will be used below,
correspond to projections of local quasiparticle operators onto
the low energy manifold. Indeed, the operators $\chi_{j,\sigma}$
($\sigma=\pm$1) carry a charge of $1/m$ and a spin of $\pm1/m$ (as
can be verified by their commutation relations with the total
charge and total spin operators). Therefore, their quantum numbers
are identical to those of a single fractional quasi-particle with
spin up or down. Moreover, the commutation relations satisfied by
$\chi_{i,\sigma}$ and for $i<j$,

\begin{align}
\chi_{i,\sigma}\chi_{j,\uparrow} & =e^{-i\frac{\pi}{m}}\chi_{j,\uparrow}\chi_{i,\sigma},\nonumber \\
\chi_{i,\sigma}\chi_{j,\downarrow} &
=e^{i\frac{\pi}{m}}\chi_{j,\downarrow}\chi_{i,\sigma},\label{eq:chi-com}
\end{align}
coincide with those of quasi-particle operators $e^{i\phi(x)\pm
i\theta(x)}$ localized at the SC/FM interfaces (for $i=j$,
$[\chi_{j,\uparrow},\chi_{j,\downarrow}]=0$ if $j$ is odd, and
satisfy
$\chi_{j,\uparrow}\chi_{j,\downarrow}=e^{2i\pi/m}\chi_{j,\downarrow}\chi_{j,\uparrow}$
if $j$ is even). Note that in our convention, for $j$ odd,
$\chi_{j,\sigma}$ corresponds to the interface between the
segments labelled by $e^{i\pi\hat{S}_{j}}$ and
$e^{\pi\hat{Q}_{j}}$, where for $j$ even, between
$e^{i\pi\hat{Q}_{j}}$ and $e^{i\pi\hat{S}_{j+1}}$, see
Fig.~\ref{fig:braiding}~a.

Therefore, the operators $\chi_{j,\sigma}$ correspond to
quasi-particle creation operators at the SC/FM interfaces,
projected onto the low-energy subspace. This conclusion is further
supported by calculating directly the matrix elements of the
microscopic quasi-particle operator between the approximate ground
states, in the limit of strong cosine potentials (see Appendix
\ref{app:MatrixElements}). This calculation reveals that the
matrix elements of the quasi-particle operators within the
low-energy subspace are proportional to those of
$\chi_{j,\sigma}$, and that the proportionality constant decays
exponentially with the distance of the quasi-particle operator
from the interface. We note that the commutation relations of
Eq.~(\ref{eq:chi-com}) appear in a one dimensional lattice model
of ``parafermions'' \cite{Fendley,Fradkin}.

\section{Topological manipulations} \label{sec:topological-manipulations}

\subsection{setup}

\begin{figure}[t]
\centerline{\includegraphics[width=1\columnwidth]{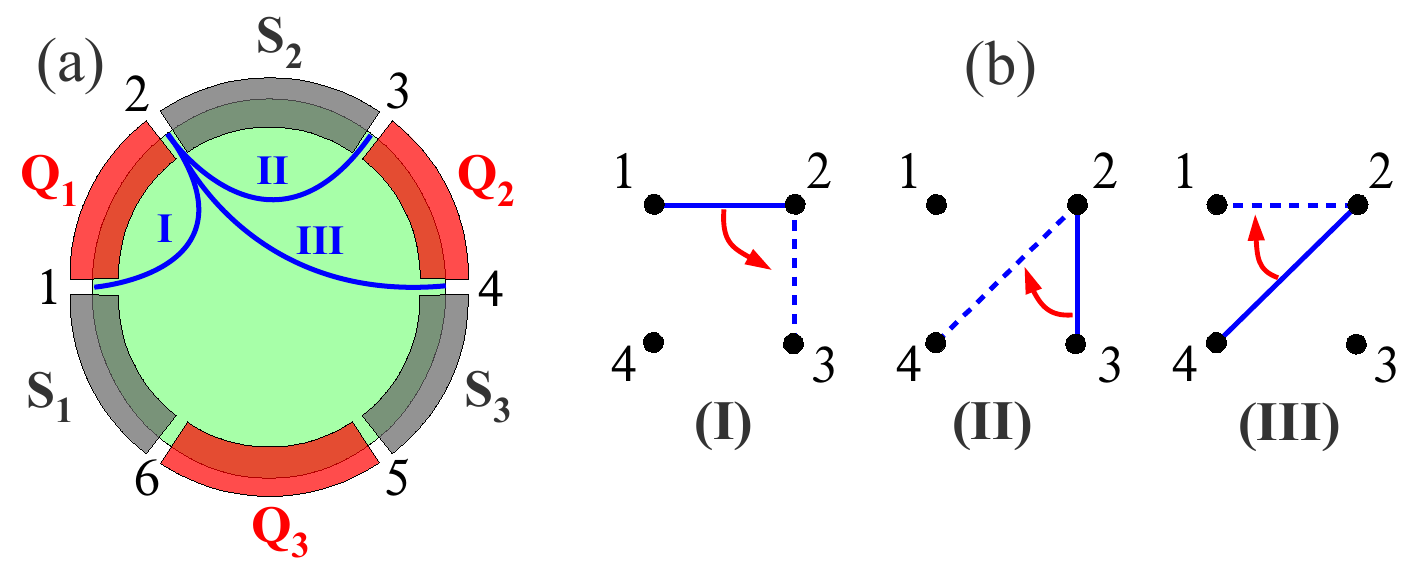}}\caption{\textbf{Braiding
process.} (a) An FTI disk with six SC/FM segments. In stages I, II
and III of the braiding process, quasi-particle tunneling
(represented by blue solid lines) is turned on between the SC/FM
interfaces. (b) Representation of the braiding procedure,
involving interfaces 1, 2, 3 and 4. In the beginning of each
stage, the two interfaces connected by a solid line are coupled;
during that stage, the bond represented by a dashed line is
adiabatically turned on, and simultaneously the solid bond is
turned off. By the end of stage III, the system returns to the
original configuration.}

\label{fig:braiding}
\end{figure}

The braiding process is facilitated by deforming the droplet
adiabatically, such that different SC/FM interfaces are brought
close to each other at every stage.  Proximity between interfaces
essentially couples them, by allowing quasi-particles to tunnel
between them. We shall assume that only one spin species can
tunnel between interfaces. The reason for this assumption will
become clear in next sections, and we shall explain how it is
manifested in realizations of the model under consideration. At
the end of the process, the droplet returns to its original form,
but the state of the system does not return to the initial state.
The adiabatic evolution corresponds to a unitary matrix acting on
the ground state manifold.

%
%
%

Below, we analyze a braid operation between nearest-neighbor
interfaces, which we label $3$ and $4$ (for later convenience).
The operation consists of three stages, which are described
pictorially in Fig.
\ref{fig:braiding}b. It begins by nucleating a new,
small, segment which is flanked by the interfaces $1$ and $2$. At
the beginning of the first stage, the small size of the new
segment means that interfaces $1$ and $2$ are coupled to each
other, and all the other interfaces are decoupled. During the
first stage, we simultaneously bring interface $3$ close to $2$,
while moving $1$ away from both $2$ and $3$, such that at the end
of the process only $2$ and $3$ are coupled to each other, while
$1$ is decoupled from them. In the second stage, interface $4$
approaches $3$, and $2$ is taken away from $3$ and $4$. In the
final stage, we couple $1$ to $2$ and decouple $4$ from $1$ and
$2$, such that the Hamiltonian returns to its initial form. In the
following, we analyze an explicit Hamiltonian path yielding this
braid operation, which is summarized in Table~\ref{tab:braiding}.
Later, we shall discuss the conditions under which the result is
independent of the specific from of the Hamiltonian path
representing the same Braid operation.

\begin{table}[t]
\caption{Summary of the braiding adiabatic trajectory (shown also in Fig. \ref{fig:braiding}b).
There are three stages, $\alpha=\mathrm{I,II,III}$, along each of
which the parameter $\lambda_{\alpha}$ varies from $0$ to $1$. The
Hamlitonian in each stage is written in the middle column, where
we use the notation
$H_{ij}=-t_{ij}\chi_{j,\uparrow}\chi_{i,\uparrow}^{\dagger}+h.c.$
($t_{ij}$ are complex parameters). The right column summarizes the
symmetry operators which commute with the Hamiltonian throughout
each stage.\label{tab:braiding}}

\begin{tabular*} {0.48\textwidth} {@{\extracolsep{\fill}} l c c}
\hline \hline Stage & Hamiltonian & Symmetries\tabularnewline
\hline I &
$\left(1-\lambda_{\mathrm{I}}\right)H_{12}$+$\lambda_{\mathrm{I}}H_{23}$
& $e^{i\pi\hat{Q}_{3}}$, $e^{i\pi\hat{S}_{3}}$\tabularnewline
\hline II &
$\left(1-\lambda_{\mathrm{II}}\right)H_{23}$+$\lambda_{\mathrm{II}}H_{24}$
& $e^{i\pi\hat{Q}_{3}}$, $e^{-i\pi\hat{S}_{1}}$\tabularnewline
\hline III &
$\left(1-\lambda_{\mathrm{III}}\right)H_{24}$+$\lambda_{\mathrm{III}}H_{12}$
& $e^{i\pi\hat{Q}_{3}}$,
$e^{-i\pi\hat{Q}_{2}}e^{i\pi\hat{S}_{3}}$\tabularnewline \hline
\hline
\end{tabular*}
\end{table}

\subsection{Ground state degeneracy}
\label{sec: degeneracy} To analyze the braiding process, we first
need to show that it does not change the ground state degeneracy.
We consider a disk with a total of $N=3$ segments of each type.
The ground state manifold, without any coupling, is $(2m)^2$ fold
degenerate. We define operators $H_{12},H_{23}$ and $H_{24}$, the
Hamiltonians at the beginning of the three stages I, II, III.
These are given by
\begin{equation}
H_{jk}=-t_{jk}\chi_{j,\uparrow}^{\vphantom{\dagger}}\chi_{k,\uparrow}^{\dagger}+h.c..\label{eq:hjk}
\end{equation}
where the $t_{jk}$ are complex amplitudes.

Consider first the initial Hamiltonian (see Table
\ref{tab:braiding}), given by

\begin{equation}
H_{12}=-t_{12}\chi_{2,\uparrow}^{\vphantom{\dagger}}\chi_{1,\uparrow}^{\dagger}+h.c.=-2\left|t_{12}\right|\cos\left(\pi\hat{Q}_{1}+\varphi_{12}\right).
\label{eq:h12}
\end{equation}
Here, $\varphi_{12}=\arg\left(t_{12}\right)$. It is convenient to
work in the basis of eigenstates of the operators $\qo$, $\qtw$,
$\qt$, and $e^{i \pi \hat{S}_{\mathrm{tot}}}$, which we label by
$\vert q_1,q_2,q_3; s \rangle$. The total charge and spin are
conserved, and we may set $\sum_j q_j=0$ and $s=0$. Then, a state
in the $\left(2m \right)^{2}$-dimensional low-energy subspace can
be labelled as $\vert q_{2},q_{3}\rangle$, where $q_{1}$ is fixed
to $q_{1}=-q_{2}-q_{3}$. The initial Hamiltonian (\ref{eq:h12}) is
diagonal in this basis, and therefore its eigen-energies can be
read off easily:
$E_{12}\left(q_{2},q_{3}\right)=-2\left|t_{12}\right|\cos\left[-\frac{\pi}{m}\left(q_{2}+q_{3}\right)+\varphi_{12}\right]$.
For generic $\varphi_{12},$ there are $2m$ ground states. Since
this FM segment is nucleated inside a SC region, its total spin is
zero, and the ground states are
\begin{equation}
|\Psi_i (q_{3})\9=\vert q_{2}=-q_{3},q_{3}\rangle,\label{eq:n3_i}
\end{equation}
labelled by a single index $q_{3}=0,\dots,2m-1$. The residual
$2m$-fold ground state degeneracy can be understood as a result of
the symmetries of the Hamiltonian. From Eq.~(\ref{eq:h12}) $\qt$
and $\st$ commute with $H_{12}$. The commutation relations between
$\qt$ and $\st$ ensures that the ground state is (at least)
$2m$-fold degenerate by the eignevalues of $\qt$.

Similar considerations can be applied in order to find the ground
state degeneracy throughout the braiding operation. The operator
$e^{i\pi\hat{Q}_{3}}$ \emph{always} commutes with the Hamiltonian,
at any stage. This can be seen easily from the fact that the
segment labelled by $\qt$ never couples to any other segment at
any stage (see Fig. \ref{fig:braiding}a). Using the definition of
the $\chi_{i\sigma}$ operators, Eq.~(\ref{eq:chi-def}), one finds
that \beq
H_{23}=-2\left|t_{23}\right|\cos\left(\pi\hat{S}_{2}+\varphi_{23}\right),
\label{eq: h23} \eeq and \beq
H_{24}=-2\left|t_{24}\right|\cos\left[\pi\left(\hat{S}_{2}+\hat{Q}_{2}\right)+\varphi_{24}\right].
\label{eq: h24} \eeq 

In each stage, $\alpha=\mathrm{I,II,III}$, there is a symmetry
operator $\Sigma_{\alpha}$ that commutes with the Hamiltonian, and
satisfies
$\Sigma_{\alpha}e^{i\pi\hat{Q}_{3}}=e^{-i\frac{\pi}{m}}e^{i\pi\hat{Q}_{3}}\Sigma_{\alpha}$.
 We specify $\Sigma_{\alpha}$ for each stage in
the right column of Table \ref{tab:braiding}, and the
aforementioned relation can be verified using Eq.~(\ref{algebra}).
This combination of symmetries dictates that \emph{every state is
at least $2m$ fold degenerate}, where each degenerate subspace can
be labelled by $q_{3}$. Assuming that the special values
$\varphi_{12},\varphi_{23}=\pi (2l+1)/(2m)$ and $\varphi_{24}=\pi
l/m$ ($l$ integer) are avoided, the ground state is exactly
$2m$-fold degenerate throughout the braiding process (the special
values for the $\varphi_{ij}$ give an additional two fold
degeneracy). Note that these conclusions hold for \emph{any}
trajectory in Hamiltonian space, as long as the appropriate
symmetries are maintained in each stage of the evolution, and the
accidental degeneracies are avoided.

\subsection{Braid matrices from Berry's phases}
\label{sub:Braiding-evolution-operator}

The evolution operator corresponding to the braid operation can
thus be represented as a block-diagonal unitary matrix, in which
each $\left(2m\right)\times\left(2m\right)$ block acts on a
separate energy subspace. We are now faced with the problem of
calculating the evolution operator in the ground state subspace.
Let us denote this operator by $\hat{U}_{34}$, corresponding to a
braiding operation of interfaces 3 and 4. The calculation of
$\hat{U}_{34}$ can be done analytically by using the symmetry
properties of Hamiltonian at each stage of the evolution.


We begin by observing that, since $\qt$ always commutes with the
Hamiltonian, $\hat{U}_{34}$ and the evolution operators for each
stage are diagonal in the the basis of $\qt$ eigenstates. In every
stage, the adiabatic evolution maps $\qt$ eigenstates between the
initial and final ground state manifolds while preserving the
eigenvalue $q_3$, and multiplies by a phase factor that may depend
on $q_{3}$. This is explicitly summarized as

\begin{equation}
\hat{U}_{\alpha}|\Psi^\alpha_{i}(q_{3})\9=\exp{\big(i\gamma_\alpha \left(q_3\right)\big)} |\Psi^\alpha_{f}(q_{3})\9.
\label{eq:gamma1}
\end{equation}
Here, $\hat{U}_\alpha$ is the evolution operator of stage
$\alpha=\mathrm{I,II,III}$, and $|\Psi^\alpha_{i(f)}(q_{3})\9$ are
the ground states of the initial (final) Hamiltonian in stage
$\alpha$, respectively, which are labelled by their $\qt$
eigenvalues. Likewise, $\gamma_{\alpha}\left(q_{3}\right)$ are the
phases accumulated in each of the stages.


In order to determine $\gamma_\alpha\left(q_3\right)$, we use the
additional symmetry operator $\Sigma_\alpha$ for each stage, as
indicated in Table~\ref{tab:braiding}. This symmetry commutes with
the Hamiltonian, and therefore also with the evolution operator
for this stage $[\Sigma_\alpha,\hat{U}_\alpha]=0$. Acting with
$\Sigma_\alpha$ on both sides of (\ref{eq:gamma1}), we get that


\begin{equation}
\hat{U}_{\alpha}\Sigma_\alpha|\Psi^\alpha_i(q_3)\9=e^{i\gamma_\alpha \left(q_3\right)}\Sigma_\alpha|\Psi^\alpha_f(q_3)\9.\label{eq:UQ}
\end{equation}

Furthermore, the relation $\qt \Sigma_\alpha = e^{i\frac{\pi}{m}}
\Sigma_\alpha \qt$ implies that the operator $\Sigma_\alpha$
advances $\qt$ by one increment, and therefore for both the
initial and final stage at each stage we have,


\begin{align}
\Sigma_\alpha |\Psi^\alpha_{i(f)}(q_3)\9&
=\exp{\left(i\delta^\alpha_{i(f)}\left(q_3\right)\right)}|\Psi^\alpha_{i(f)}(q_3+1)\9,\label{eq:expQ}
\end{align}
where $\delta^\alpha_{i(f)}\left(q_3\right)$ are phases which
depends on gauge choices for the different eigenstates, to be
determined below. Inserting (\ref{eq:expQ}) into (\ref{eq:UQ}), we
get the recursion relation

\begin{equation}
\gamma_{\alpha}\left(q_3+1\right)=\gamma_{\mathrm{\alpha}}\left(q_3\right)+\delta^\alpha_{f}\left(q_3\right)-\delta^\alpha_{i}\left(q_3\right).
\label{eq:gamma-I}
\end{equation}
Note that while the phase accumulation at each point along the
path depends on gauge choices, the total Berry phase accumulated
along a cycle does not. It is convenient to choose a
\textit{continuous} gauge, for which the total Berry's phases are
given by
\begin{equation}
\hat{U}_{34}|\Psi^\alpha_{i}(q_3)\9=\exp{\big(i \sum_\alpha \gamma_\alpha \left(q_3\right)\big)} |\Psi^\alpha_{i}(q_3)\9.\label{eq:gamma}
\end{equation}
A continuous gauge requires
$|\Psi_f^\alpha(n_3)\9=|\Psi_i^{\alpha+1}(n_3)\9$. Therefore, the
values of the phases $\delta^\alpha_{i(f)}$ depend only on three
gauge choices. These are the gauge choices eigenstates of the
Hamiltonians $H_{12}$, $H_{23}$, and $H_{24}$, which constitute
the initial Hamiltonian at the beginning of stages I-III, as well
as the final Hamiltonian for stage III.


Making the necessary gauge choice, allows us to solve
Eq.~(\ref{eq:gamma-I}) for $\gamma_\alpha(n_3)$, yielding the
total Berry phase (the details of the calculation are given in
Appendix \ref{sec:braiding-app})
\begin{equation}
\gamma\left(q_3\right)=\frac{\pi}{2m}(q_3-k)^{2}.\label{eq:gamma-tot}
\end{equation}
The integer $k$ depends on the choice for the phases
$\varphi_{ij}$. Recall that the Hamiltonians $H_{ij}$,
Eqs.~(\ref{eq:h12}),(\ref{eq: h23}--\ref{eq: h24}), have an
additional degeneracy for a discrete choice of the $\varphi_{ij}$.
Any two choices for the $\varphi_{ij}$ that can be deformed to
each other without crossing a degeneracy point yield the same $k$.

The evolution operator for the braiding path can be written
explicitly by its application on the eigenstates of the
Hamiltonian in the beginning of the cycle,
$\hat{U}^{(k)}_{34}|\Psi_i
(q_3)\9=e^{i\frac{\pi}{2m}(q_3-k)^{2}}|\Psi_i (q_3)\9$. Since by
Eq.~(\ref{eq:n3_i}), the ground states of the initial Hamiltonian
satisfy $q_{2}=-q_3$, this can be written in a basis-independent
form in terms of the operator $\qtw$. Loosely speaking,
$\hat{U}_{34}$ can be written as
\begin{equation}
\hat{U}^{(k)}_{34}=\exp\left(\frac{i\pi
m}{2}\left(\hat{Q}_{2}-\frac{k}{m}\right)^{2}\right).\label{eq:U34-1}
\end{equation}
Alternatively, using the identity\cite{gauss}
$e^{i\frac{\pi}{2m}q^{2}}=\sqrt{\frac{1}{2m}}\sum_{p=0}^{2m-1}e^{i\frac{\pi}{m}\left(pq-\frac{p^{2}}{2}\right)+i\frac{\pi}{4}}$,
one can write
\begin{equation}
\hat{U}^{(k)}_{34}=\sqrt{\frac{1}{2m}}\sum_{p=0}^{2m-1}e^{-\frac{i\pi}{2m}(p+k)^{2}+i\frac{\pi}{4}}\left(e^{i\pi\hat{Q}_{2}}\right)^{p}.\label{eq:U34-2}
\end{equation}

In the case $m=1$, $\hat{U}_{34}$ reduces to the braiding rule of
Ising anyons \cite{Ivanov,ReadGreen,NayakWilczek}. 

\begin{figure}[t]
\includegraphics[width=1\columnwidth]{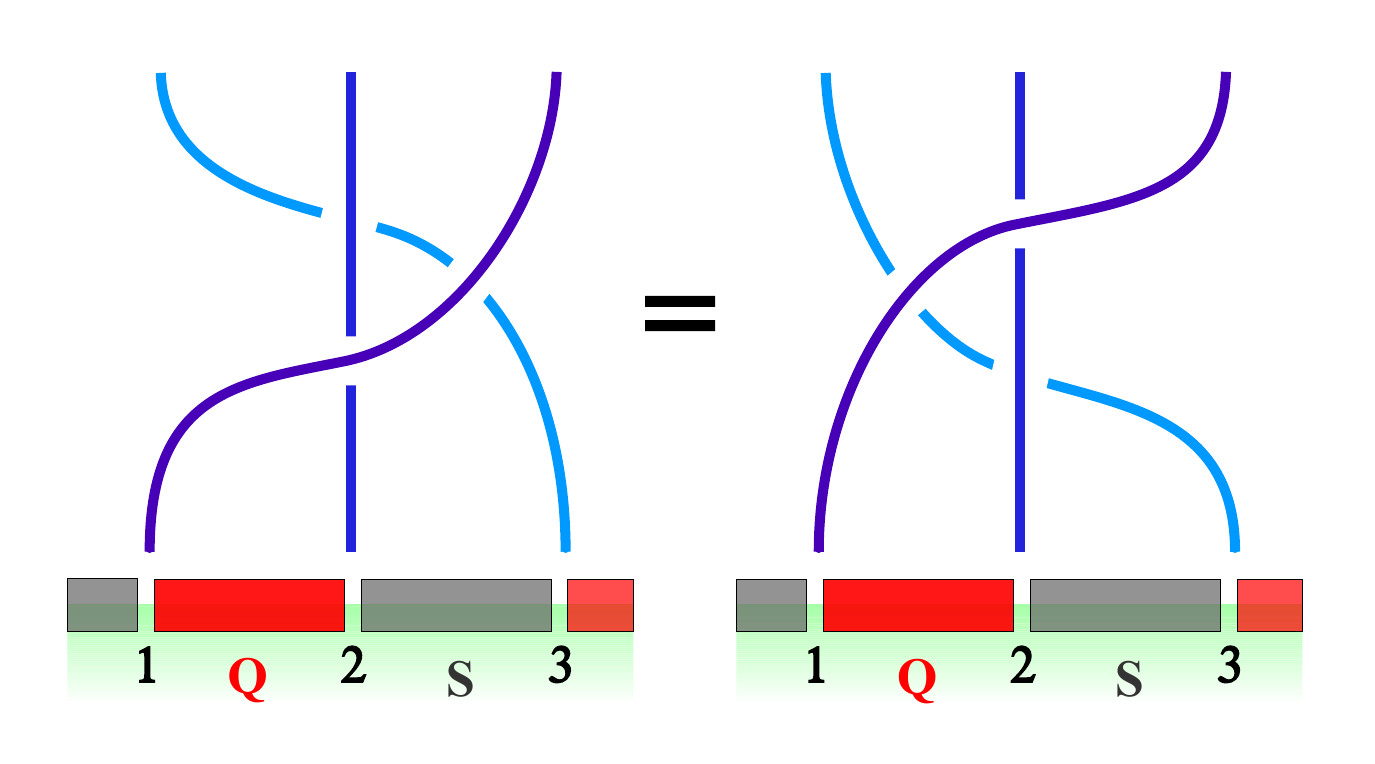}\caption{\textbf{Diagrammatic representation of the Yang-Baxter
equations} (Eq. \ref{eq:YB1}). Three interfaces 1,2,3 are braided
in two distinct sequences. The Yang-Baxter equations state that
the results of these two sequences of braiding operations are the
same.} \label{fig:YB}
\end{figure}

Following a similar procedure, one can construct the operator
representing the exchange of any pair of neighboring interfaces:
$\hat{U}^{(k)}_{2j-1,2j}=e^{\frac{i\pi
m}{2}\left(\hat{S}_{j}-k/m\right)^{2}}$,
$\hat{U}^{(k)}_{2j,2j+1}=e^{\frac{i\pi
m}{2}\left(\hat{Q}_{j+1}-k/m\right)^{2}}$. In order for these
operations to form a representation of the braid group, it is
necessary and sufficient that they satisfy
\begin{align}
\left[\hat{U}^{(k_i)}_{i,i+1},\hat{U}^{(k_j)}_{j,j+1}\right] & =0\,\,\,(\left|i-j\right|>1),\label{eq:YB-1}\\
\hat{U}^{(k_1)}_{j,j+1}\hat{U}^{(k_2)}_{j+1,j+2}\hat{U}^{(k_1)}_{j,j+1} &
=\hat{U}^{(k_2)}_{j+1,j+2}\hat{U}^{(k_1)}_{j,j+1}\hat{U}^{(k_2)}_{j+1,j+2}.\label{eq:YB-2}
\end{align}
Equation~(\ref{eq:YB-1}) clearly holds because the spin or charge
operators of non-nearest neighbor segments commute. Using
(\ref{eq:U34-2}), it is not difficult to show that (\ref{eq:YB-2})
holds as well (see Appendix~\ref{app:yang-baxter}.
Eq.~(\ref{eq:YB-2})  is depicted in Fig.~\ref{fig:YB}). Therefore,
$\hat{U}_{i,i+1}$ form a representation of the braid group. In
that respect, our system exhibits a form of non-abelian
statistics. By combining a sequence of nearest-neighbor exchanges,
an exchange operation of arbitrarily far segments can be defined.

In any physical realization, we do not expect to control the
precise form of the Hamiltonian in each stage. It is therefore
important to discuss the extent to which the result of the
braiding process depends on the details of the Hamiltonian along
the path. We argue that the braiding is ``topological'', in the
sense that it is, to a large degree, \textit{independent} of these
precise details.

To see this, one needs to note that the braiding unitary matrix
was derived above without referring to the precise adiabatic path
in Hamiltonian space. All we used were the \emph{symmetry
properties} of the Hamiltonian in each stage (Table
\ref{tab:braiding}). These symmetries do not depend on the precise
details of the intermediate Hamiltonian, but only on the overall
configuration, e.g., which interfaces are allowed to couple in
each stage.

In Appendix \ref{sec: independece}, we state more formally the
conditions under which the result of the braiding is independent
of details. Special care must be taken in stage III of the
braiding, in which \emph{quasi-particles of only one spin
species}, e.g. spin up, must be allowed to tunnel between
interfaces 2 and 4. We elaborate on the significance of this
requirement and the ways to meet it in the various physical
realizations in Appendix \ref{sec: symmetries}.

\section{Braiding and topological spin of boundary anyons}
\label{sec: TS}

In the previous section, we derived the unitary matrix
representing braid operations by an explicit calculation. In the
following, we shall try to shed light on the physical picture
behind these representations. To do so, we show that the results
of the previous section can be derived almost painlessly, just by
assuming that the representation of the braids have properties
which are analogous to those of anyons in two dimensions. The
first and most basic assumption, is very natural: there exists a
topological operation in the system which corresponds to a braid
of two interfaces, in that the unitary matrices representing this
operation obey the Yang-Baxter equation.

The operations we consider braid two neighboring interfaces, but
\textit{do not} change the total charge (spin) in the segment
between them. This results from the general form of the braid
operations - to exchange two interfaces flanking a SC (FM)
segment, we use couplings to an auxiliary segment, of the same
type. Therefore, charge (spin) can only be exchanged with the
auxiliary segment. Since the auxiliary segment has zero charge
(spin) at the beginning and end of the operation, the charge of
the main segment cannot change by the operation. Indeed, this can
be seen explicitly in the analysis presented in the previous
section. As a result, the unitary matrix representing the braid
operation is diagonal with respect to the charge (spin) of the
segment.

The derivation now proceeds by considering a property of anyons
called the topological spin (TS).  In two dimensions, the
topological spin gives the phase acquired by a $2\pi$ rotation of
an anyon. For fermions and bosons, the topological spin is the
familiar $-1$ and $+1$ respectively (corresponding to half-odd or
integer spins). There is a close connection between the braid
matrix for anyons and their topological spin. In two dimensions,
these relations have been considered by various authors
\cite{KitaevHex, PreskillNotes}. The system under consideration is one
dimensional, and therefore seemingly does not allow a $2\pi$
``rotation'' of a particle. However, as we shall explain below,
the TS of a particle can be defined in our system using the
relations of the TS to the braid matrix. We shall then see how to
use these relations to derive the possible unitary representations
of the braid operations in the system at point.

In our one dimensional system, we consider the TS of two different
kinds of objects (particles) - interfaces, which we denote by $X$,
and the charge (or spin) of a segment, which we shall label by
$q=0,1,..2m-1$. In what follows, we need to know how to compose,
or \textit{fuse} different objects in our system. As we saw above,
two interfaces yield a quantum number $\exp(i \pi Q)$ ($\exp(i \pi
S)$) which is the total charge (spin) in the segment between them,
respectively. Suppose we consider two neighboring SC segments with
quantum numbers $\exp(i \pi Q_1)$ and $\exp(i \pi Q_2)$, and we
``fuse'' them by shrinking the FM region which lies between them.
This results in tunneling of fractional quasi-particles between
the two SC regions, and energetically favors a specific value for
$\exp(i
\pi S)$ in the FM region. The two SC segments are for all purposes
one, where clearly, in the absence of other couplings,
$\exp\left(i \pi (Q_1 + Q_2)\right)$ remains a good quantum
number. This therefore suggests the following fusion rules \bea X
\times X &=& 0 + 1 + ... + 2m-1
\nonumber\\ q_1 \times q_2 &=& (q_1 + q_2) \mod 2m \label{eq:
fusion1} \eea We note that the labelling $q_1=0,1...,2m-1$ does
not depend on the gauge choices in the definition of the operators
$\exp(i\ \pi Q)$. Equation~(\ref{eq: fusion1}) suggests that the
labelling can be defined by the addition law for charges, in which
each type of charge plays a different role. Indeed, this addition
rule has a measurable physical content which does not depend on
any gauge choices.

In two dimensional theories of anyons, it is convenient to think
about particles moving in the two dimensional plane, and consider
topological properties of their world lines (such as braiding). In
this paper, we have defined braiding by considering trajectories
in Hamiltonian space. In the following, we represent these
Hamiltonian trajectories as world-lines of the respective
``particles'' involved, keeping in mind that they do not
correspond to motion of objects in real space.

\begin{figure}[t]
\centerline{\includegraphics[width=1\columnwidth]{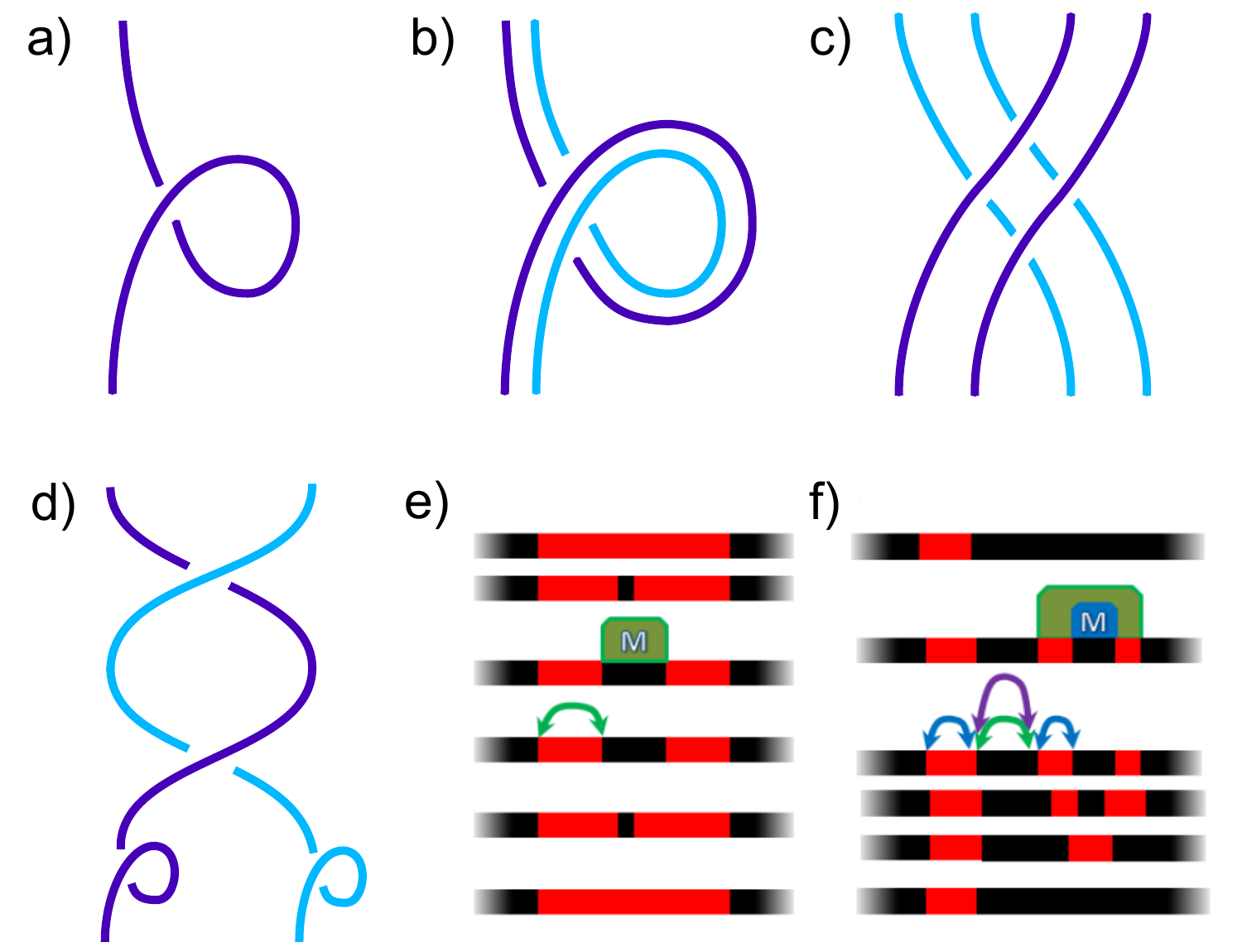}}
\caption{\textbf{Topological Spin.} The process illustrated in (a)
defines the TS, where the cup $\bigcup$ corresponds to creation of
a particle -anti particle pair (two $X$'s fusing to $q=0$ or two
charges with $q_1+q_2=0$), and the cap $\bigcap$ corresponds to
projection on zero total charge.  (b) TS of a composite object.
When both lines are labelled by $X$ and fuse to charge $q$ , the
phase acquired is $\theta_q$. The $4$-fold crossing, which is
magnified in (c), is assumed to result in an exchange of two
$q$'s. When the lines are labelled by $q_1$ and $q_2$, the phase
acquired is $\theta_{q_1+q_2 \mod 2m}$. Importantly, the TS of a
composite is equal to the phase accumulated in the process
appearing in (d). This equality results from using the Yang-Baxter
equation and the definition of the TS. (e) The evolution of the
different segments in the process defining $\theta_X$. The green
arrow corresponds to braiding, while the ``M'' corresponds to
projection on zero charge. (f) The process defining $\theta_q$.
The braids are ordered by color, green, blue and finally purple.
The projections are on zero spin for the magnet segment
intersecting the two SC segments, and zero total charge for these
two segments.} \label{fig: TS}
\end{figure}

We are now ready to define the TS in our system. In short, the TS
of a particle is a phase factor associated to the world line
appearing in Fig~\ref{fig: TS}(a). For interfaces, it is
concretely defined by the phase acquired by the system by the
following sequence of operations, as illustrated in Fig~\ref{fig:
TS}(e): (i) nucleation of a segment to the right (by convention)
of the interface $X_1$ (note that the notation $X_1$ corresponds
to particle $X$ at coordinate $r_1$). The total spin or charge of
this segment is zero (the nucleation does not add total charge to
the system). The couplings between $X_2$ and $X_3$ flanking the
new segment is taken to zero, increasing the ground state
degeneracy by a factor of $2m$. (ii) A right handed braid
operation is performed between $X_1$ and $X_2$. (iii) The total
charge $q$ of the segment between $X_2$ and $X_3$ is measured, and
we consider (post-select) only the outcomes corresponding to
\textit{zero} charge. Therefore, the system ends up in the same
state (no charges have been changed anywhere in the system), up to
a phase factor. Importantly, this phase factor does not depend on
the state of the system, since the operation does not change the
total charge in the segment of $X_1$ and $X_2$ (see Appendix
\ref{sec: TS app} for a more detailed discussion). We can
therefore define this phase factor as $\theta_X$, the topological
spin of particle type $X$.

In order to define $\theta_q$, the topological spin of a charge
$q$, we first need to define the operation corresponding to an
exchange of two charges. Consider the sequence of $4$ right handed
exchanges of the interfaces, as in Fig.~\ref{fig: TS} (c). The
figure suggests that this sequence should yield an exchange of the
fusion charges $q_1$ and $q_2$ of the two pairs of $X$ particles,
as would indeed be the case for anyons in two dimensions (see
Appendix.~\ref{sec: TS app} for more details). Our second
assumption is that this is indeed the case. Since by Eq.~(\ref{eq:
fusion1}) there is only one fusion channel for the $q_i$'s,  the
state is multiplied by a phase factor which depends only on $q_1$
and $q_2$ - the charges $q_i$ are \textit{abelian}. It is
straightforward to check that exchanges of charges satisfy the
Yang-Baxter equation.

The operations defining $\theta_q$ are illustrated in
Figs~\ref{fig: TS}(b) and (f). We consider, say, a SC segment in
an eigenstate of $\exp(i\pi Q_1)$ to which we associate a particle
type $q_1$. In step (i), we nucleate two segments to the right of
$q_1$. This step is actually done in two substeps: first, we
nucleate a SC segment to the right of the segment $q_1$, and then
we nucleate a FM segment which separates this segment into two
segments, with charges $\exp\left(i \pi
\left(Q_2 + Q_3\right)\right)=1$. Note that the spin of the middle
FM segment is also zero, $\exp\left(i \pi S_2\right)=1$. We now
perform a braid between the segments labelled $q_1$ and $q_2$.
Next, we measure the charges $\exp\left(i\pi \left(Q_2 +
Q_3\right)\right)$ and $\exp(i \pi S)$, post-selecting their
values to be equal to $1$. The charge $q_1$ in the original
segment is therefore unchanged by this sequence of operations, as
well as any other charges used to label the original state of the
system. As before, the state of the system acquires a phase, which
only depends on the charge $q_1$.

We now use an important relation between the TS of a composite to
a double braid of its two components. This is a result of an
equality between Fig~\ref{fig: TS} (b) and (d), which can be
derived by using the Yang-Baxter equation and the definition for
the TS. The equality between Fig.~\ref{fig: TS} (b) and (d), with
both incoming and outgoing lines labelled by $X$, yields \beq
\theta_q=\theta_X^2  U^2(q) \label{eq: top spin} \eeq where $U(q)$
corresponds to the unitary matrix representing an exchange of two
interfaces whose fusion charge is $q$.  We shall not attempt to
calculate $\theta_X$. However, by calculating the topological spin
$\theta_q$ of the composite, we can get, using Eq.~(\ref{eq: top
spin}), the square of the sought after braid matrix, up to a
global phase.

In two dimensional theories of anyons, the TS of a composite of
two particles is equal to the TS of the particle they fuse into.
This can be understood by noting that the equality between
Fig~\ref{fig: TS} (b) and (d) guarantees that in two dimensions,
the TS of the composite includes both the intrinsic spin of the
two particles and their relative angular momentum (see
Appendix~\ref{sec: TS app} for more details). We now assume the
same holds in our one dimensional system - the TS of the $q_1$,
$q_2$ composite is equal to the TS of combined $q=q_1+q_2
\mod 2m$ charge.  The TS of a composite of two charges $q_1$ and $q_2$ is
just the process described in Fig.~\ref{fig: TS}(b), with the
outgoing lines are labelled with $q_1$ and $q_2$.

Next, we note that for consistency, the charge $q=0$ must
correspond to trivial TS, $\theta_0=1$. We take the TS of the
elementary charge $q=1$ as a parameter, $\theta_{1}=e^{i\alpha}$,
to be determined later. Importantly, note that Fig.~\ref{fig:
TS}(a) implies that an exchange of two segments with $q=1$ leads
to the phase factor $\theta_1$. Using the composition rule in
Eq.~(\ref{eq: fusion1}), $\theta_q$ should be equal to the TS of a
composite of $q$ unit charges, a process where we encounter $q^2$
exchanges of these elementary charges. This gives for the
topological spin \beq \theta_{q \!\!\!\mod 2m}=\exp{\left(i \alpha
q^2\right)}. \label{eq: top spin q} \eeq The $\mathbb{Z}_{2m}$
structure above appears since fusing $2m$ elementary particles
results in the trivial $q=0$ charge, and requires that \beq \alpha
= \frac{
\pi p}{2m}, \qquad p \in \mathbb{Z}. \label{eq: alpha} \eeq

Taking the square root of Eq.~(\ref{eq: top spin q}) we arrive at
\beq U(q)=(-1)^{f(q)}\exp{\left(i \frac{\alpha}{2} q^2\right)}.
\eeq where $f(q)$ is an arbitrary, integer valued function. To
determine $\alpha$ and $f(q)$ we appeal to the Yang-Baxter
equation, Eq.~(\ref{eq:YB-2}). We find numerically that solutions
are possible only for even $p$'s in Eq.~(\ref{eq: alpha}), where
sign function
$(-1)^{f(q)}$ 
can be of the form $e^{i\pi (n_1q^2+n_2)}$ with integer $n_1,n_2$.
Therefore, the extra $(-1)$ signs can always be absorbed in the
definition of $\alpha$, up to an overall sign. The final result is
therefore \beq U(q)=\exp{\left(i \frac{\pi n}{2m} q^2 \right)},
\qquad n \in \mathbb{Z}. \eeq

As noted in Sec.~\ref{sec:physical-picture}, the value $n=1$ is
realized by quasi-particle tunnelling of a single spin species,
while $n=m$ is realized by electron tunnelling. The other
representations of the braid group, which are given by
$U^{(k)}(q)=\exp{\left(i \frac{\pi n}{2m} (q-k)^2 \right)}$ can be
obtained from the above considerations by adding more  particles
types. Indeed, we see that \beq
U^{(k)}(q)\left(U^{(k-1)}(q)\right)^{\dag}=e^{-i\frac{\pi
n}{2m}(2k-1)} e^{i\frac{\pi n}{m} q} \eeq is, up to a global
phase, a topological operation which is equivalent to taking a
vortex $n$ times around a SC segment, or changing the orientation
of the FM direction $n$ times in the $x-y$ plane by $2\pi$.

\section{Quantum information processing}
\label{sec:quantum-computing}

In order to assess the suitability of our system for topologically
protected quantum computation, it is instructive to examine the
structure of the resulting non-abelian theory. Below, we show that
the representation of the braid group realized in our system is a
direct product of an Ising anyonic theory times a novel
representation of dimension $m$. We argue that braiding operations
alone are not sufficient to realize universal topologically
protected quantum computation, in agreement with the general
argument for models with anyons of quantum dimension which is a
square root of an integer\cite{Rowell2009}.

The unitary matrix that describes the braiding operation of two
interfaces at the ends of a superconducting segment,
Eq.~(\ref{eq:U34-1}), depends on the charge in the superconducting
segment. The charge can be written as $Q=q/m$ where $q$ can be
uniquely expressed as $q=m q_\sigma + 2 q_\upsilon$, with
$q_\sigma=0,1$ and $q_\upsilon=0,\dots,m-1$. Inserting this
expression into Eq.~(\ref{eq:U34-1}), and assuming for simplicity
$k=0$, we get \beq \hat{U} =e^{\frac{i\pi}{2}q_\sigma^2}
e^{\frac{2i\pi}{m} q_\upsilon^2}. \label{eq:U_tensor} \eeq
Therefore, we see that if we write the Hilbert space $\mathcal{H}$
as a tensor product $\mathcal{H_\sigma} \otimes
\mathcal{H_\upsilon}$, such that the states are written as $\vert
q \rangle = \vert q_\sigma \rangle \otimes \vert q_\upsilon
\rangle$, the braiding matrix decomposes into a tensor product
$\hat{U}_\sigma \otimes \hat{U}_\upsilon$. Here, $\hat{U}_\sigma$
and $\hat{U}_\upsilon$ are $2\times 2$ and $m\times m$ matrices,
given by the first and second terms on the right hand side of
Eq.~(\ref{eq:U_tensor}), respectively. A similar decomposition
holds for a braid operation acting on a ferromagnetic segment. In
this respect, we see that the $2m$-dimensional representation of
the braid group given by Eq.~(\ref{eq:U34-1}) is \emph{reducible}:
it decomposes into a two-dimensional representation, which is
nothing but the representation formed by Ising anyons $\sigma$,
times an $m$-dimensional representation corresponding to braiding
of a non-Ising object $\upsilon$.

This decomposition gives insights into the class of unitary
transformations that can be realized using braiding of interfaces,
and hence their suitability for quantum computation. We now argue
that by using braiding operations alone, the system studied in
this paper does not allow for topologically protected universal
quantum computation. Ising anyons are known not to provide
universality for quantum computation \cite{RMP, Freedman,
FreedmanLarsen}. Due to the tensor product structure of the
topologically protected operations, it is sufficient to consider
the fractional part, corresponding to $\mathcal{H_\upsilon}$. The
braid representations acting within this subspace
preserves\cite{us-future-work} a generalization of the Pauli group
to q-dits of dimension $m$. Therefore, the braid operations can be
simulated on a classical computer, and are not universal.

Conceptually, universality could be achieved by adding an
entangling operation between the Ising and the fractional parts. A
braiding operation would then produce an effective phase gate that
would provide the missing ingredient to make the Ising part
universal. However, at present we do not know whether it is
possible to realize such an operation in a topologically protected
way. Moreover, topologically protected measurements of charges
\cite{Hassler} (and spins) cannot achieve such entanglement, since
similarly to the braiding, they can also be shown to be of a
tensor product form.

\section{Concluding remarks} \label{sec:conclusions}

In this work, we have described a physical route for utilizing
proximity coupling to superconductors in order to realize a
species of non-abelian anyons, which goes beyond the Majorana
fermion paradigm. The essential ingredients of the proposed system
are a pair of counter-propagating edge modes of a Laughlin
fractional quantum Hall state, proximity-coupled to a
superconductor. As we saw, there are several possible realizations
of such a system. One could start from a ``fractional topological
insulator'' whose edges are coupled to an array of superconductors
and ferromagnets. In the absence of any known realization of a
fractional topological insulator phase (as of today), one could
get by starting from ``ordinary'' Laughlin fractional quantum Hall
state whose edges are coupled to a superconductor. The fractional
quantum Hall state in graphene might be a promising candidate for
realizing such systems, since the magnetic fields needed for
observing it are much lower than the fields needed in
semiconductor heterostructure devices. 

An experimentally accessible signature of the fractionalized
Majorana modes is a fractional Josephson effect, which should
exhibit a component of $4m\pi$ periodicity (analogously to the
$4\pi$ periodicity predicted for topological superconductors with
Majorana edge modes\cite{FuPRB,Liang}). In addition, it might be
possible to observe topological pumping of fractional charge by
controlling the relative phase of the superconducting regions.

More broadly speaking, the system we describe here is an example
of how gapping out the edge state of a fractionalized
two-dimensional phase can realize a topological phase which
supports new types of non-abelian particles, not present in the
original two-dimensional theory. In our example, the underlying
Laughlin fractional quantum Hall state supports quasiparticles
with a fractional charge and fractionalized \emph{abelian}
statistics; the resulting gapped theory on the edge, however,
realizes \emph{non-abelian} quasiparticles. Moreover, the
resulting non-abelian theory on the edge is shown to go beyond the
well-known Majorana (Ising) framework.

This may seem contradictory to the general
arguments\cite{Lukasz1,Lukasz2, Turner, Chen, Schuch} indicating
that gapped one-dimensional systems with no symmetry other than
fermion parity conservation support only two distinct topological
phases, a trivial phase and a non-trivial phase, with an odd
number of Majorana modes at the interface between them. The reason
our system avoids this exhaustive classification is that it is
not, strictly speaking, one-dimensional; the edge states of
fractional quantum Hall states can never be realized as degrees of
freedom of an isolated one-dimensional system. This is reflected,
for example, in the fact that the theory contains ``local'' (from
the edge perspective) operators which satisfy fractional
statistics, which is not possible in any one-dimensional system
made of fermions and bosons.

It would be interesting to pursue this idea further, by examining
gapped states which are realized by gapping out edge modes of
topological phases. This may serve as a route to discovering new
classes of topological phases with non-abelian excitations. For
example, more complicated \cite{Halperin} quantum Hall states or
higher dimensional fractional topological
insulators\cite{Levin3D,Swingle,Maciejko} may be interesting
candidates for such investigations.

\begin{acknowledgments}
We thank Maissam Barkeshli, Lukasz Fidkowsky, Bert Halperin,
Alexei Kitaev, Chetan Nayak and John Preskill for useful
discussion. E. B. was supported by the NSF under grants
DMR-0757145 and DMR-0705472. A. S. thanks the US-Israel Binational
Science Foundation, the Minerva foundation, and Microsoft Station
Q for financial support. N. H. L. and G. R. acknowledges funding
provided by the Institute for Quantum Information and Matter, an
NSF Physics Frontiers Center with support of the Gordon and Betty
Moore Foundation, and DARPA. N. H. L. was also supported by the
David and Lucile Packard Foundation.
\end{acknowledgments}

\emph{Note added.-} During the course of working on this
manuscript, we became aware that a similar idea is being pursued
by David Clarke, Jason Alicea, and Kirill
Shtengel\cite{Clarke2012}. After this manuscript was submitted,
two papers on related subjects\cite{Cheng2012,Vaezi2012} have
appeared.

\appendix
\section{Appendix: Matrix elements of the quasi-particle operator}
\label{app:MatrixElements}
 In this Appendix, we describe an
explicit calculation of the matrix element of a quasi-particle
operator between different states in the ground state manifold. We
show that the matrix element is finite if the quasi-particle is
located sufficiently close to an interface between a
superconducting and a ferromagnetic segment. The matrix element
decays exponentially with the distance from the interface.

\subsection{Model}

Let us consider a system composed of one superconducting segment,
extending from $x=0$ to $x=L$, between two long ferromagnetic
segments at $x<0$ and $x>L$. For simplicity, we assume that the
gap in the ferromagnetic segments is very large, such that charge
fluctuations are completely quenched outside the superconductor.
The Hamiltonian for $0\le x\le L$ is

\begin{align}
H & =\int_{0}^{L}dx\left\{ \frac{m}{2\pi}u\left[\left(\partial_{x}\phi\right)^{2}+\left(\partial_{x}\theta\right)^{2}\right]-g_{S}\cos\left(2m\phi\right)\right\} \nonumber \\
\label{eq:Hsc}
\end{align}
supplemented by the boundary condition

\begin{equation}
\partial_{x}\phi\left(x=0,L\right)=0,
\end{equation}
which accounts for the fact that the current at the edges of the
superconductor is identically zero, due to the large gap in the
ferromagnetic regions. We are assuming that the coupling $g_{S}$
is large enough such that the field $\phi$ is pinned to the
vicinity of the minima of the cosine potential,
$\phi\approx\frac{\pi}{m}l$, where $l$ is an integer. Deep in the
superconducting phase, one can expand the cosine potential up to
second order around one of the minima, obtaining the effective
Hamiltonian

\begin{align}
H_{\mathrm{eff}} & =\int_{0}^{L}dx\left\{ \frac{m}{2\pi}u\left[\left(\partial_{x}\phi\right)^{2}+\left(\partial_{x}\theta\right)^{2}\right]+\frac{g}{2}\left(\phi-\frac{\pi l}{m}\right)^{2}\right\} ,\nonumber \\
\label{eq:Hquad}
\end{align}
where $g\equiv\left(2m\right)^{2}g_{S}$.

The Hamiltonian (\ref{eq:Hquad}) is quadratic, and can be
diagonalized using the following mode expansion:

\begin{equation}
\phi\left(x\right)=\frac{\pi}{m}l+\hat{\phi}_{0}+i\frac{1}{\sqrt{m}}\sum_{k=1}^{\infty}\sqrt{\frac{1}{k}}\cos\left[\frac{\pi k}{L}x\right]\left(a_{k}-a_{k}^{\dagger}\right),\label{eq:ModePhi}
\end{equation}

\begin{equation}
\theta\left(x\right)=\theta\left(0\right)+\frac{\pi\hat{n}}{mL}x+\frac{1}{\sqrt{m}}\sum_{k=1}^{\infty}\sqrt{\frac{1}{k}}\sin\left[\frac{\pi k}{L}x\right]\left(a_{k}+a_{k}^{\dagger}\right).\label{eq:ModeTheta}
\end{equation}

Here, we have introduced the ladder operators $a_{k}$,
$k=1,2,\dots,$ satisfying
$\left[a_{k},a_{k'}^{\dagger}\right]=\delta_{k,k'}$ and
$\left[a_{k},a_{k'}\right]=0$. $\hat{\phi}_{0}$ and $\hat{n}$ are
the average phase and the charge of the superconducting segment.
These variables are canonical conjugates, satisfying
$\left[\hat{\phi}_{0},\hat{n}\right]=i$. Note that
$\left[\theta\left(0\right),H_{\mathrm{eff}}\right]=0$, therefore
$\theta\left(0\right)$ can be replaced by a c-number:
\begin{equation}
\theta\left(0\right)=\frac{\pi}{m}p,
\end{equation}
with integer $p$, where we have assumed that these values minimize
the (infinite) cosine potential on the ferromagnetic side $x<0$.
Using Eq.~(\ref{eq:ModePhi}),(\ref{eq:ModeTheta}), one can
reproduce the commutation relation
$\left[\phi\left(x\right),\theta\left(x'\right)\right]=i\frac{\pi}{m}\Theta\left(x'-x\right)$.

Inserting the mode expansions into the Hamiltonian
(\ref{eq:Hquad}), we get

\begin{align}
H_{\mathrm{\mathrm{eff}}} & =\frac{\pi u}{2mL}\hat{n}^{2}+\frac{gL}{2}\hat{\phi}_{0}^{2}+\sum_{k=1}^{\infty}\left(\frac{\pi uk}{2L}+\frac{gL}{4mk}\right)\left[a_{k}^{\dagger}a_{k}+a_{k}a_{k}^{\dagger}\right]\nonumber \\
 & -\sum_{k=1}^{\infty}\frac{gL}{4mk}\left(a_{k}^{2}+a_{k}^{\dagger2}\right).
\end{align}
This Hamiltonian is diagonalized by a Bogoliubov transformation of
the form
\begin{equation}
a_{k}=\alpha_{k}b_{k}+\beta_{k}b_{k}^{\dagger},
\end{equation}
where
$\alpha_{k}=\sqrt{\frac{1}{2}\left(\frac{A_{k}}{E_{k}}+1\right)}$,
$\beta_{k}=\sqrt{\frac{1}{2}\left(\frac{A_{k}}{E_{k}}-1\right)}$,
and $E_{k}=\sqrt{A_{k}^{2}-B_{k}^{2}}$, expressed via
$A_{\ensuremath{k}}=\frac{\pi uk}{L}+\frac{gL}{2mk}$ and
$B_{k}=\frac{gL}{2mk}$. The $\hat{n}$, $\hat{\phi}_{0}$ part of
$H_{\mathrm{eff}}$ is diagonalized by introducing ladder operators
$\eta$, $\eta^{\dagger}$ such that

\begin{align}
\hat{\phi} & =\left(\frac{gmL^{2}}{\pi u}\right)^{-1/4}\frac{\hat{\eta}+\hat{\eta}^{\dagger}}{\sqrt{2}},\nonumber \\
\hat{n} & =\left(\frac{gmL^{2}}{\pi u}\right)^{1/4}\frac{\hat{\eta}-\hat{\eta}^{\dagger}}{i\sqrt{2}}.
\end{align}
The diagonal form of $H_{\mathrm{eff}}$ (up to constants) is

\begin{equation}
H_{\mathrm{SC}}=E_{0}\hat{\eta}^{\dagger}\hat{\eta}+\sum_{k=1}^{\infty}E_{k}b_{k}^{\dagger}b_{k},
\end{equation}
where $E_{0}=\sqrt{\pi gu/m}$.

\subsection{Computation of the matrix elements}

Next, we calculate matrix elements of a quasi-particle creation
operator between states in the ground state manifold, which we
index by the average values of $\theta$ and $\phi$ on either side
of the $x=0$ interface. A diagonal matrix element has the form

\begin{equation}
A_{l,p}\left(x\right)=\langle\psi_{l,p}\vert
e^{i\left[\phi\left(x\right)+\theta\left(x\right)\right]}\vert\psi_{l,p}\rangle,
\end{equation}
where $\vert\psi_{l,p}\rangle=\vert\pi l/m,\pi p/m\rangle$ is a
ground state in which $\phi\left(x>0\right)$ and
$\theta\left(x<0\right)$ are localized near $\frac{\pi}{m}l$ and
$\frac{\pi}{m}p$, respectively. Note that these two variables
\emph{commute}, and therefore they can be localized
simultaneously. To evaluate $A_{l,p}\left(x\right)$, we use the
identity
\begin{equation}
\langle e^{\hat{O}}\rangle=e^{\langle\hat{O}\rangle+\frac{1}{2}\left(\langle\hat{O}^{2}\rangle-\langle\hat{O}\rangle^{2}\right)},
\end{equation}
valid for any operator $\hat{O}$ which is at most linear in
creation and annihilation operators. Substituting
$\hat{O}=i\left[\phi\left(x\right)+\theta\left(x\right)\right]$,
the expectation values in the exponent can be computed using the
mode expansions~(\ref{eq:ModePhi}), (\ref{eq:ModeTheta}). The
computaion is lengthy but straightforward, giving

\begin{equation}
A_{m,n}\left(x\right)=e^{i\frac{\pi}{m}\left(l+p\right)-\frac{1}{2}\left(F_{\phi}\left(x\right)+F_{\theta}\left(x\right)\right)},\label{eq:Amn}
\end{equation}
where

\begin{align}
F_{\phi}\left(x\right) & =\frac{1}{2L}\left(\frac{mg}{\pi u}\right)^{-1/2}\nonumber \\
 & +\sum_{k=1}^{\infty}\frac{\pi u}{mL}\frac{\cos^{2}\left[\frac{\pi k}{L}x\right]e^{-\frac{\alpha\pi k}{L}}}{\sqrt{\left(\frac{\pi uk}{L}\right)^{2}+\pi\nu gu}}\label{eq:Phi_sq}
\end{align}

\begin{align}
F_{\theta}\left(x\right) & =\left(\frac{mg}{\pi u}\right)^{1/2}\frac{\left(\pi x/m\right)^{2}}{2L}\nonumber \\
 & +\sum_{k=1}^{\infty}\frac{\nu}{k}\frac{\left(\frac{\pi uk}{L}+\frac{gL}{mk}\right)\sin^{2}\left(\frac{\pi k}{L}x\right)e^{-\frac{\alpha\pi k}{L}}}{\sqrt{\left(\frac{\pi uk}{L}\right)^{2}+\frac{\pi}{m}gu}}.\label{eq:Theta_sq}
\end{align}
Here, we have introduced exponential damping factors of the form
$e^{-\frac{\alpha k}{L}}$, where $\alpha$ is a short-distance
cutoff, to suppress ultraviolet singularities.
$F_{\phi}\left(x\right)=\langle\phi^{2}\left(x\right)\rangle-\left\langle
\phi\left(x\right)\right\rangle ^{2}$, and similarly for
$F_{\theta}$. (Expectation values of the form $\left\langle
\phi\left(x\right)\theta\left(x\right)\right\rangle $ vanish.)

We now analyze the asymptotic behavior of $F_{\phi}$ and
$F_{\theta}$ for $\xi\ll x\ll L$, where we have defined the
correlation length as $\xi=\sqrt{mu/\left(\pi g\right)}$. In the
limit $L\rightarrow\infty$, the sums over $k$ in Eq.
(\ref{eq:Phi_sq}),(\ref{eq:Theta_sq}) can be replaced by integrals
over $q\equiv k/L$. Then, the long-distance asymptotic behavior of
$F_{\phi}$ and $F_{\theta}$ is easily extracted:

\begin{equation}
F_{\phi}\left(x\right)\sim\frac{1}{2m}\log\left(\frac{\xi}{\alpha}\right),
\end{equation}

\begin{equation}
F_{\theta}\left(x\right)\sim\frac{1}{2m\xi}\left[-\frac{1}{2}\alpha\log\frac{\alpha^{2}+4x^{2}}{\alpha^{2}}+ix\log\left(\frac{\alpha-2ix}{\alpha+2ix}\right)\right].
\end{equation}
Inserting these expressions into (\ref{eq:Amn}) gives

\begin{equation}
A_{l,p}\left(x\right)\sim
e^{i\frac{\pi}{m}\left(l+p\right)-\frac{\pi}{2m\xi}x}.
\end{equation}
Therefore, the diagonal matrix element of the quasi-particle
operator in the ground state $\vert\psi_{l,p}\rangle$ decays
exponentially with the distance from the interface. In a very
similar way, one can show that the matrix element of the
quasi-particle operator between two ground states with different
$\{l,p\}$ vanishes in the limit $L\rightarrow\infty$. It is
therefore natural to identify the operators $\chi_{i,\sigma}$
introduced in Sec.~\ref{sec:domain-wall-operators} as the
projection onto the ground state subspace of the quasi-particle
operators $e^{i\left(\phi(x)\pm\theta(x)\right)}$ acting at the
interface.

\section{Calculation of the braid matrix}
\label{sec:braiding-app} In order to complete the calculation of
the unitary matrix corresponding to the braiding operation of two
interfaces, we first need to obtain the ground states of the
Hamiltonians $H_{12}$, $H_{23}$ and $H_{24}$, making the necessary
gauge choices. As in Sec.~\ref{sec:topological-manipulations}, we
use the basis of eigenstates of the operators $\exp(i\pi
\hat{Q}_j)$, $j=1,2,3$. We work in the sector $\prod_j \exp(i\pi
\hat{Q}_j)=1$. A state in this sector can be labelled as $\vert
q_2,q_3\rangle$, where $\exp(\frac{i\pi}{m} q_j)$ is the
eigenvalue of $\exp(i\pi \hat{Q}_j)$ and $q_1 = -q_2-q_3$.

The ground state of $H_{12}=-2|t_{12}|\cos(\pi
\hat{Q}_1+\varphi_{12})$ is given by \beq
|\psi_{i}^\mathrm{I}(q_3)\rangle=|\psi_{f}^{\mathrm{III}}(q_3)\rangle=|q_2=-q_3+k_{\mathrm{I}},
q_3\rangle, \label{eq: psi12 appendix} \eeq where the integer
$k_\mathrm{I}$ is determined by $\varphi_{12}$ according to \beq
\frac{\pi}{m}(k_\mathrm{I}-\frac{1}{2})<\varphi_{12}<\frac{\pi}{m}(k_\mathrm{I}+\frac{1}{2}).
\eeq The ground state is $2m$-fold degenerate, corresponding to
the $2m$ possible values of $q_3$. Equation~(\ref{eq: psi12
appendix}) includes an explicit gauge choice for the ground
states. Note that for
$\varphi_{12}=\frac{\pi}{m}(k_\mathrm{I}+\frac{1}{2})$, the ground
state degeneracy increases to $4m$. We therefore assume that these
values of $\varphi_{12}$ are avoided.

The Hamiltonian in the beginning of the second stage, $H_{23}$,
can be written in the basis of $q_2$ eigenstates as
\begin{equation}
H_{23}=-\left|t_{23}\right|\sum_{q_{2}=0}^{2m-1}e^{i\varphi_{23}}\vert
q_{2}\rangle\langle q_{2}+1\vert+h.c..\label{eq:H24}
\end{equation}
The above form can be derived from the relation
$\qtw\stw=e^{i\pi/m}\stw\qtw$, i.e. $\stw$ is a raising operator
for $\qtw$. The Hamiltonian (\ref{eq:H24}) can be thought of as an
effective tight-binding model on a periodic ring of length $2m$
with complex hopping amplitudes. Note that the total effective
flux through the ring is given by
\begin{equation}
\Phi^{23}_{\mathrm{eff}}=2m \varphi_{23}.
\end{equation}
Importantly, note that when $\Phi^{23}_{\mathrm{eff}}=\pi$, the
ground state of $H_{23}$ is doubly degenerate. These are
degeneracy points that we assume are avoided in the braiding
process.

The ground state for a particle on a ring with flux
$\Phi^{23}_{\mathrm{eff}}$ is simply a plane wave, \beq
\vert\psi^{\mathrm{I}}_f(q_3)\rangle=\vert\psi^{\mathrm{II}}_i(
q_3)\rangle=\sum_{n=0}^{2m-1}e^{i\frac{\pi}{m} n
k_{\mathrm{II}}}|q_2=-q_3+n, q_3\rangle, \label{eq: psi23} \eeq
where $k_{\mathrm{II}}$ is the closest integer to
$-(m\varphi_{23})/\pi$. Note that again, a gauge choice for the
overall phase of the states has been made in Eq.~(\ref{eq:
psi23}).

The phases $\delta^\mathrm{I}_{i,f}(q_3)$, defined in
Eq.~(\ref{eq:expQ}) of the main text, are determined by operating
with the symmetry operator $\Sigma_\mathrm{I} = \exp(\pi i
\hat{S}_3)$ on the ground states of the initial $(H_{12})$ or the
final $(H_{23})$ Hamiltonian, Eqs.~(\ref{eq: psi12
appendix}),(\ref{eq: psi23}), respectively. In the gauge we have
chosen, this gives
\begin{equation}
\delta^I_{f}\left(q_{3}\right)=\delta^I_{i}\left(q_{3}\right)=0.
\qquad  \label{eq:delta-zero}
\end{equation}
Therefore, the recursion relation for $\gamma_\alpha (q_3)$,
\begin{equation}
\gamma_{\alpha}\left(q_3+1\right)=\gamma_{\mathrm{\alpha}}\left(q_3\right)+\delta^\alpha_{f}\left(q_3\right)-\delta^\alpha_{i}\left(q_3\right),\label{eq:gamma-I-app}
\end{equation}
leads to \beq
\gamma_{\mathrm{I}}\left(q_{3}+1\right)=\gamma_{\mathrm{I}}(q_3).
\eeq Since only the differences between the Berry phases of
different states matter, $\gamma_{\mathrm{I}}\left(q_{3}\right)$
are defined up to an arbitrary overall phase. This overall phase
can be chosen such that, for stage I, $\gamma_\mathrm{I}(q_3)=0$.



The Hamiltonian at the end of stage II, $H_{24}$, can be written
as
\begin{align}
H_{24} & =-\left|t_{24}\right|e^{i\varphi_{24}}\chi^{\vphantom{\dagger}}_{4\uparrow}\chi_{2\uparrow}^{\dagger}+h.c\nonumber \\
 & =-\left|t_{24}\right|e^{i\varphi_{24}}e^{i\pi\hat{S}_{2}}e^{i\pi\hat{Q}_{2}}+h.c.
\end{align}
In the second line, we have used the explicit form of
$\chi_{2,4\uparrow}$ in terms of the spin and charge operators.
Writing the Hamiltonian in the basis of eigenstates of
$e^{i\pi\hat{Q}_{2}}$, we get
\begin{equation}
H_{24}=-\left|t_{24}\right|\sum_{q_{2}=0}^{2m-1}e^{i\varphi_{24}+i\frac{\pi}{m}q_{2}}\vert
q_{2}+1\rangle\langle q_{2}\vert+h.c.\label{eq:H24}
\end{equation}

In order to diagonalize $H_{24}$, we perform a gauge
transformation to a new basis $\vert \xi_{q_2}\rangle$ defined as
\begin{equation}
\vert
q{}_{2}\rangle=e^{-i\frac{\pi}{2m}q_{2}^{2}}\vert\xi_{q_{2}}\rangle.\label{eq:gauge}
\end{equation}
This transformation is designed such that, in the new basis, the
phases of the hopping amplitudes are uniform. The Hamiltonian
takes the form
\begin{equation}
H_{24}=-\left|t_{24}\right|\sum_{q_{2}=0}^{2m}e^{i\left(\varphi_{24}-\frac{\pi}{2m}\right)}
\vert\xi_{q_{2}+1}\rangle\langle\xi_{q_{2}}\vert+h.c.
\end{equation}
which is easily diagonalized in the basis of plane waves. One can
verify that, defining
$\vert\tilde{\xi}_{p}\rangle=\frac{1}{\sqrt{2m}}\sum_{q_{2}=0}^{2m-1}e^{i\frac{\pi}{m}p
q_{2}}\vert\xi_{q_{2}}\rangle$,
\begin{equation}
H_{24}\vert\tilde{\xi}_{p}\rangle=-2\left|t_{24}\right|\cos\left[\frac{\pi}{m}(p+\frac{1}{2})-\varphi_{24}\right]\vert\tilde{\xi}_{p}\rangle.
\end{equation}
Therefore, for
$\frac{\pi}{m}k_{\mathrm{III}}<\varphi_{24}<\frac{\pi}{m}\left(k_{\mathrm{III}}+1\right)$
where $k_{\mathrm{III}}$ is an integer, we get that the ground
state occurs for $p=k_{\mathrm{III}}\mathrm{mod}\left(2m\right)$.
Using Eq. (\ref{eq:gauge}), one can express the ground state in
terms of the original basis states:
\begin{equation}
\vert\psi^{\mathrm{II}}_f(q_3)\rangle=\vert\psi^{\mathrm{III}}_i(q_3)\rangle
=\frac{e^{-i\frac{\pi}{2m}k_{\mathrm{III}}^{2}}}{\sqrt{2m}}
\sum_{q_{2}=0}^{2m-1}e^{i\frac{\pi}{2m}\left(q_{2}+k_{\mathrm{III}}\right)^{2}}\vert
q{}_{2}\rangle. \label{eq:psi24}
\end{equation}

Applying the symmetry operator
$\Sigma_{\mathrm{II}}=e^{-i\pi\hat{S}_{1}}$ to both sides of
(\ref{eq: psi12 appendix}),(\ref{eq:psi24}), we get \beq
\delta^{\mathrm{II}}_{i}\left(q_{3}\right)=-\frac{\pi}{m}
k_{\mathrm{II}},\qquad \delta^{\mathrm{II}}_{f}\left(q_{3}\right)=
0. \eeq Therefore, solving the recursion relation (Eq.
\ref{eq:gamma-I-app}) and choosing a gauge such that
$\gamma_{\mathrm{II}}(q_3=0)=0$, we get \beq
\gamma_{\mathrm{II}}(q_3) = \frac{\pi}{m}
k_{\mathrm{II}}q_3.\label{eq:gamma-II}\eeq

For the last stage of the evolution, Applying
$\Sigma_{\mathrm{III}}=e^{-i\pi\hat{Q}_{2}}e^{i\pi\hat{S}_{3}}$ to
both sides of Eqs.~(\ref{eq:psi24}),(\ref{eq: psi12 appendix}), we
get \beq
\delta^{\mathrm{III}}_{i}\left(q_{3}\right)=\frac{\pi}{m}(k_{\mathrm{III}}+\frac{1}{2}),
\qquad
\delta^{\mathrm{III}}_{f}\left(q_{3}\right)=\frac{\pi}{m}(q_3-k_{\mathrm{I}}+1).
\label{eq: delta III} \eeq Inserting this into Eq.
\ref{eq:gamma-I-app} and solving for $\gamma_{\mathrm{III}}(q_3)$,
we obtain \beq \gamma_{\mathrm{III}}(q_3) =
\frac{\pi}{2m}(q_3-k_{\mathrm{I}}-k_{\mathrm{III}})^2.
\label{eq:gamma-III}\eeq

The total Berry phase
$\gamma=\gamma_{\mathrm{I}}+\gamma_{\mathrm{II}}+\gamma_{\mathrm{III}}$,
up to an unimportant overall phase, is
\begin{equation}
\gamma\left(q_{3}\right)=\frac{\pi}{2m}(q_{3}-k)^{2}.\label{eq:gamma-tot}
\end{equation}
Here, $k = k_{\mathrm{I}}-k_{\mathrm{II}}+k_{\mathrm{III}}$. Note
that while $\gamma_\alpha$ depend on our various gauge choices for
the basis of the eigenstates of $H_{12}$, $H_{23}$, and $H_{24}$,
the Berry phases of the entire path (Eq. \ref{eq:gamma-tot}) does
not depend on these gauge
choices. 

\section{Topological Protection of the Braid Operations}
\label{sec: protection app}

\subsection{Independence of microscopic details}
\label{sec: independece}

In any physical realization, one would not be able to control the
precise form of the Hamiltonian in each stage of the braid
process. It is therefore important to discuss to what extent the
result of the braiding depends on the details of the Hamiltonian
along the path. Below, we argue that the braiding is
``topological'', in the sense that it is independent of these
precise details.

Let us begin by noting that the evolution operator describing the
full braiding process depends only on:
\begin{enumerate}
\item The initial and final Hamiltonians at each stage; \item The
symmetries of the Hamiltonian at each stage; \item The fact that
the ground state degeneracy throughout the process is fixed, such
that the evolution can be considered adiabatic.
\end{enumerate}

One can see that the precise details of the time-dependent
Hamiltonian during the braiding process are unimportant for
our derivation of the evolution operator in Sec.~\ref{sec:topological-manipulations}. 
Note that we have never used the exact form of the Hamiltonian
during the path to determine the evolution operator.

\label{sec: general}
\begin{figure}[t]
\centerline{\includegraphics[width=1.0\columnwidth]{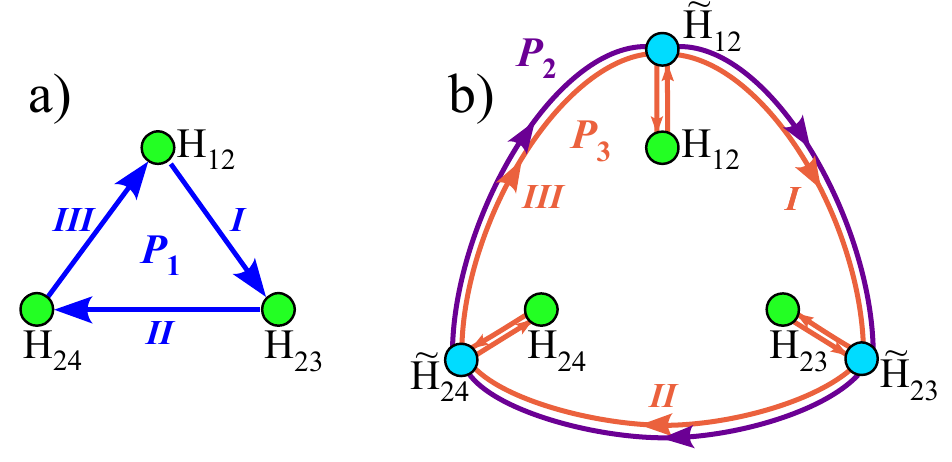}}\caption{\textbf{Braiding
paths in Hamiltonian space.} (a) Path $P_1$, for which we compute
the braiding adiabatic evolution operator explicitly. (b) A
different path $P_2$, whose Hamiltonians at the intermediate
stages are assumed to be adiabatically connectable to those of
$P_1$. $P_3$ is a path equivalent to $P_2$, in which each
intermediate Hamiltonian of $P_2$ evolves to the corresponding
Hamiltonian of $P_1$ and then back.} \label{fig:path}
\end{figure}

In order to make this argument more formal, let us define $P_1$ as
the closed path in Hamiltonian space, $H_{12}\rightarrow
H_{23}\rightarrow H_{24}\rightarrow H_{12}$, for which we computed
the evolution operator. ($P_1$ is summarized in Table
\ref{tab:braiding}.) Suppose that we replace $P_1$ by a different,
``realistic'' path $P_2$, defined as
$\tilde{H}_{12}\rightarrow\tilde{H}_{23}\rightarrow\tilde{H}_{24}\rightarrow\tilde{H}_{12}$,
which has the same symmetries as those of the original trajectory
in each stage (Table \ref{tab:braiding}). $P_1$ and $P_2$ are
represented in Fig. \ref{fig:path} a and b, respectively. We
assume further that the Hamiltonian at the end of every stage of
$P_2$ is adiabatically connectable to that of the original path
$P_1$, e.g. $\tilde{H}_{12}$ and $H_{12}$ are adiabatically
connectable, etc. We argue that the adiabatic evolution associated
with $P_2$ is unitarily equivalent to that of $P_1$. To show this,
consider the modified path $P_3$ shown in Fig. \ref{fig:path}:
\begin{align}
H_{12} & \rightarrow\tilde{H}_{12}\rightarrow\tilde{H}_{23}\rightarrow H_{23}\rightarrow\tilde{H}_{23}\nonumber \\
\rightarrow & \tilde{H}_{24}\rightarrow H_{24}\rightarrow\tilde{H}_{24}\rightarrow\tilde{H}_{12}\rightarrow H_{12}.\label{eq:path}
\end{align}
Clearly, Eq.~(\ref{eq:path}) can be viewed as a \emph{deformed}
version of the original trajectory $P_1$, in which the
intermediate Hamiltonian during each stage is deformed relative to
the original trajectory of Table \ref{tab:braiding}. Since the
intermediate Hamiltonian in every stage of trajectory
(\ref{eq:path}) has the same symmetries as those of the original
trajectory, the analysis outlined in the previous section shows
that the evolution operator representing the overall trajectory
(\ref{eq:path}) is $e^{i\omega}\hat{U}_{34}$, where $\omega$ is
global phase factor.

On the other hand, we can consider $P_3$ starting and ending with
$\tilde{H}_{12}$, as the ``realistic'' trajectory $P_2$. Since for
example, the step $\tilde{H}_{23}\rightarrow H_{23}$ is ``undone''
by the next step $H_{23}\rightarrow\tilde{H}_{23}$, $P_2$ is
unitarily related to the $P_3$ by $e^{i \omega} V \hat{U}_{34}
V^{\dag}$, where $V$ represents the evolution from
$\tilde{H}_{12}$ to $H_{12}$. In essence, the matrix $V$ relates
the eigenstates of the ``realistic'' initial Hamiltonian
$\tilde{H}_{12}$, to those of $H_{12}$. We conclude that the
adiabatic evolutions corresponding to paths $P_1$ and $P_2$ are
physically equivalent, and therefore the braiding operation is
robust to changes in the path in Hamiltonian space, as long as the
conditions $1-3$ listed above are met.

\subsection{Symmetries of the Hamiltonian during the braiding process}
\label{sec: symmetries}

Next, we discuss the symmetry requirements in every stage in more
physical terms. The topological stability of the braiding
operation depends crucially on the symmetries of the Hamiltonian
throughout the different stages of the braiding operation. 
We now argue that these symmetry properties are largely
independent of the microscopic details of the Hamiltonian in each
stage. This is since the definition of the braid operation only
contains information regarding which interfaces are brought in
proximity at each stage. For instance, any Hamiltonian trajectory
corresponding to this braid operation only couples interfaces $1$,
$2$ and $3$ during stage I (see Fig. \ref{fig:braiding}). Any such
Hamiltonian necessarily commutes with $e^{i\pi\hat{S}_{3}}$ and
$e^{i\pi\hat{Q}_{3}}$, \emph{independently of its microscopic
details}. For example, adding terms such as higher powers of
$\chi_{2,\uparrow}\chi_{1,\uparrow}^{\dagger}$, retains these
symmetries. Likewise, terms representing direct coupling between
interfaces $1$ and $3$, such as powers of
$\chi_{3,\uparrow}\chi_{1,\uparrow}^{\dagger}$ can be added, as
long as they are absent at the beginning and end of stage I, when
interfaces $1$ and $3$ are far apart.


A similar statement can be made for stage II: as long as
interfaces $5$, $6$, and $1$ remain decoupled throughout the
evolution, the Hamiltonian necessarily maintains the same
symmetries as those in in Table~\ref{tab:braiding}, regardless of
the microscopic details of the process.


The symmetry requirement in stage III requires more care. At this
stage, interfaces $1$,$2$ and $4$ are coupled. Crucially, we note
that the commutation relations in Eq.~(\ref{eq:chi-com}) give
$\left[\chi_{i\downarrow},\chi_{j\uparrow}\chi^{\dag}_{k\uparrow}\right]=0$
for any $j,k\neq i$. Therefore, as long as we allow tunneling of
only spin-up particles between interfaces $1$,$2$ and $4$, we are
assured that $\chi_{3,\downarrow}$ commutes with the Hamiltonian.
It follows that
$\chi^{\phantom{\dag}}_{5,\downarrow}\chi^{\dag}_{3,\downarrow}=e^{-i\pi\hat{Q}_{2}}e^{i\pi\hat{S}_{3}}$,
the symmetry operator required in stage III, also commutes with
the Hamiltonian. Again, this symmetry would be maintained
\emph{independently} of the exact form of the Hamiltonian, as long
as it obeys the above restriction. The physical reason behind this
symmetry is clarified by noting that transferring $n$ up-spin
quasi-particles to interface $4$ (from either $1$ or $2$) changes
$e^{i\pi\hat{Q}_{2}}\rightarrow e^{i\pi(\hat{Q}_{2}+\frac{n}{m})}$
and $e^{i\pi\hat{S}_{3}}\rightarrow
e^{i\pi(\hat{S}_{3}+\frac{n}{m})}$, leaving
$e^{-i\pi\hat{Q}_{2}}e^{i\pi\hat{S}_{3}}$ invariant.


We now see why it is crucial, in order to allow for the braiding
operation, to have only one species of quasi-particles tunnelling
between interfaces.  If quasi-particles of both spins are allowed
to tunnel, $e^{-i\pi\hat{Q}_{2}}e^{i\pi\hat{S}_{3}}$ would cease
to be a good symmetry - in fact, the symmetry of the Hamiltonain
is lowered, and the ground state degeneracy is reduced from
$\left(2m\right)$ to just $2$, violating the adiabatically of the
braiding process. This is a special property of the fractional
($m>1$) case; for $m=1$, there is no difference between up and
down quasiparticle tunneling.

We note that the restriction to single species tunneling was
unnecessary in stages I and II, which retain the same symmetries
even when both spin species are allowed to tunnel. Moreover, if we
allowed only \emph{spin down} quasi-particles  to tunnel between
$2$ and $4$, there would be an alternative symmetry operator
$e^{i\pi\hat{Q}_{2}}e^{i\pi\hat{S}_{3}}$ at stage III. The braid
operation with this type of coupling would yield a unitary
operator of the same form found in
Sec~\ref{sec:topological-manipulations}.

\begin{figure}[t]
\centerline{\includegraphics[width=1\columnwidth]{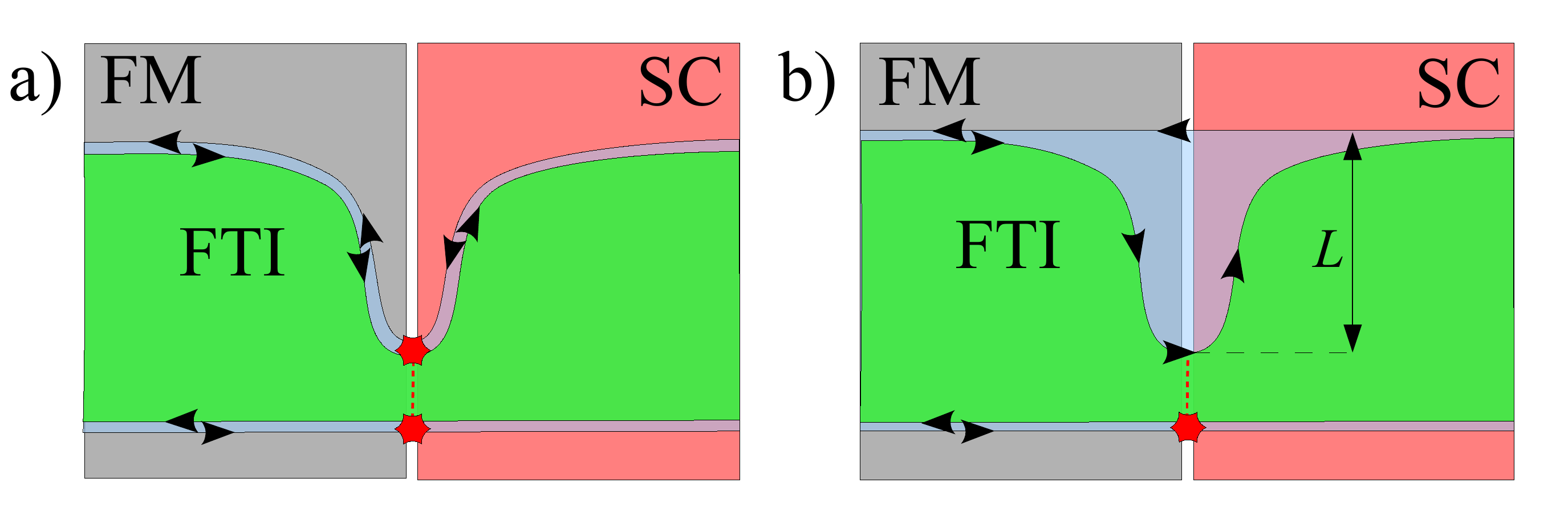}}\caption{\textbf{Constrictions
in an FTI bar.} We view the FTI as a combination of two FQH
droplets with opposite spins and opposite filling fractions. The
arrows show the direction of propagation of the edge states of the
two FQH bars. The red stars mark the position of the SC/FM
interfaces on the edges. (a) Constriction created by a gate
potential, which acts on both spin species. In this case,
quasi-particles of both spin species can tunnel across the
constriction. (b) Applying a local Zeeman field splits the spin up
and spin down edge states, such that only one spin species can
tunnel across the constriction.}

\label{fig:bridges}
\end{figure}

The restriction of single spin species tunnelling can be met in
the different realizations of the model analyzed above. First,
consider the realization using a fractional quantum Hall liquid,
in which an insulating trench separates two counter propagating
edge states (Fig. \ref{fig:system}b). In this realization, the
labels spin up and down indicate whether the quasi-particles are
on the inner or outer edge respectively. When we deform the system
in order to put interfaces in proximity, we must specify whether
this deformation shrinks the inner or outer droplet of the quantum
Hall liquid. Suppose we shrink the inner droplet. Then
quasi-particle tunnelling between interfaces proceeds through the
inner droplet. Therefore in this case, quasi-particles can only
tunnel from the inner edge at one interface location to the inner
edge at another.  Only electron tunneling processes are allowed
between the interfaces on the outer edge. However, electron
tunnelling from the outer and inner edge are equivalent, as these
are related by cooper-pair tunnelling or ``spin flip'' operators.
To conclude, in this realization, choosing whether to deform the
inner or outer quantum Hall droplet selects which spin species of
quasi-particles are allowed to tunnel between interfaces.


Let us now consider the realization of the system on the edge of a
Fractional Topological Insulator. Suppose that one can apply
either ordinary gate potentials, or Zeeman fields in the $z$
direction (by coupling to a nearby ferromagnet polarized along
$z$), which act as opposite gate potentials for the two spin
species. Then there are two ways of coupling two interfaces,
depicted in Fig. \ref{fig:bridges}. One can either create a
constriction in both spin species by applying an appropriate gate
voltage (Fig. \ref{fig:bridges}a), which allows quasi-particle
tunnelling of both spin species between the interfaces across the
constriction, or create a constriction for one spin species only,
e.g. spin up (Fig. \ref{fig:bridges}b), in which case only that
spin species tunnels. Note that in the latter case, we have split
the spin up and spin down edge states into two counter-propagating
edge modes, which become gapless. However, if the length of the
split region is $L$, there still is a finite-size gap of the order
of $v_{F}/L$, where $v_{F}$ is the Fermi velocity on the edge. The
tunnelling of quasi-particles of spin up across the constriction,
on the other hand, is enhanced by a factor of the order of $
\exp\left(L/\xi\right)$ relative to that of spin down, where $\xi$
is the correlation length in the bulk. Therefore, the tunnelling
of spin up quasi-particles can, in principle, be enhanced
parametrically without reducing the gap considerably.

\section{Yang-Baxter equations}
\label{app:yang-baxter}

Here, we verify that the unitaries representing braiding of two
neighboring interfaces by tunneling of quasi-particles satisfy the
Yang-Baxter equations. Imagine that we start from three
consecutive interfaces, 1, 2, and 3, shown in Fig. \ref{fig:YB}.
The segment between 1 and 2 is a superconducting (SC) segment, and
the segment between 2 and 3 is a ferromagnetic (FM) segment.
$e^{i\pi\hat{Q}}$ and $e^{i\pi\hat{S}}$ are the charge and spin
operators acting on the SC and FM segments, respectively. In terms
of these operators, one can express the unitary matrices that
correspond to braiding (1,2) and (2,3):

\begin{align}
U_{12} & =e^{i\frac{\pi m}{2}\hat{Q}^{2}}=\frac{1}{\sqrt{2m}}\sum_{k=0}^{2m-1}e^{-i\frac{\pi}{2m}k^{2}+i\frac{\pi}{4}}e^{i\pi k\hat{Q}},\nonumber \\
U_{23} & =e^{i\frac{\pi
m}{2}\hat{S}^{2}}=\frac{1}{\sqrt{2m}}\sum_{k=0}^{2m-1}e^{-i\frac{\pi}{2m}k^{2}+i\frac{\pi}{4}}e^{i\pi
k\hat{S}}.\label{eq:Us}
\end{align}
Here, we have used the expansion of the braiding matrices in terms
of the spin and charge operators and their harmonics.

The Yang-Baxter equations state that
\begin{equation}
U_{12}U_{23}U_{12}=U_{23}U_{12}U_{23}.\label{eq:YB1}
\end{equation}
This relation can be understood pictorially, as shown in Fig.
\ref{fig:YB}. Inserting Eqs. (\ref{eq:Us}) into the left hand side
of (\ref{eq:YB1}), and using
$e^{i\pi\hat{Q}}e^{i\pi\hat{S}}=e^{-i\frac{\pi}{m}}e^{i\pi\hat{S}}e^{i\pi\hat{Q}}$,
we get

\begin{align}
U_{12}U_{23}U_{12} & =\sum_{k_{1},k_{2},k_{3}}\frac{e^{-i\frac{\pi}{2m}\left(k_{1}^{2}+k_{2}^{2}+k_{3}^{2}\right)+i\frac{3\pi}{4}}}{\left(2m\right)^{\frac{3}{2}}}e^{i\pi k_{1}\hat{Q}}e^{i\pi k_{2}\hat{S}}e^{i\pi k_{3}\hat{Q}}\nonumber \\
 & =\sum_{k_{1},k_{2},k_{3}}\frac{e^{-i\frac{\pi}{2m}\left[\left(k_{1}-k_{2}\right)^{2}+k_{3}^{2}\right]+i\frac{3\pi}{4}}}{\left(2m\right)^{\frac{3}{2}}}e^{i\pi k_{2}\hat{S}}e^{i\pi\left(k_{1}+k_{3}\right)\hat{Q}}.
\end{align}
The sums over $k_{1,2,3}$ run from $0$ to $2m-1$. Changing
variables $k_{2}\rightarrow k_{1}+k_{2}$, and $k_{3}\rightarrow
k_{3}-k_{1}$,

\begin{align}
U_{12}U_{23}U_{12} & =\sum_{k_{1},k_{2},k_{3}}\frac{e^{-i\frac{\pi}{2m}\left[k_{2}^{2}+\left(k_{3}-k_{1}\right)^{2}\right]+i\frac{3\pi}{4}}}{\left(2m\right)^{\frac{3}{2}}}e^{i\pi\left(k_{1}+k_{2}\right)\hat{S}}e^{i\pi k_{3}\hat{Q}}\nonumber \\
 & =\sum_{k_{1},k_{2},k_{3}}\frac{e^{-i\frac{\pi}{2m}\left(k_{1}^{2}+k_{2}^{2}+k_{3}^{2}\right)+i\frac{3\pi}{4}}}{\left(2m\right)^{\frac{3}{2}}}e^{i\pi k_{2}\hat{S}}e^{i\pi k_{3}\hat{Q}}e^{i\pi k_{1}\hat{S}}\nonumber \\
 & =e^{i\frac{\pi m}{2}\hat{S}^{2}}e^{i\frac{\pi m}{2}\hat{Q}^{2}}e^{i\frac{\pi m}{2}\hat{S}^{2}}=U_{23}U_{12}U_{23}.
\end{align}
In the second line we have commuted $e^{i\pi k_{1}\hat{S}}$ with
$e^{i\pi k_{3}\hat{Q}}$. This establishes the Yang-Baxter equation
for the braiding matrices (\ref{eq:Us}). Conjugating
Eq.~(\ref{eq:YB1}) by $e^{i\pi\hat{Q}_L k_1}e^{-i\pi\hat{S}_R
k_2}$, where $e^{i\pi\hat{Q}_L}$ correspond to the charge in the
SC segment on the left of $e^{i\pi\hat{Q}}$, and
$e^{i\pi\hat{S}_R}$ to the spin in the FM segment on the right of
$e^{i\pi\hat{S}}$, we get Eq.~(\ref{eq:YB-2}).

\section{More on the Braiding and Topological Spin of Boundary
anyons}
\label{sec: TS app}

In Sec~\ref{sec: TS}, we have derived the representation of the
braid group using an analogy to braiding properties of anyons in
two dimensions. The derivation proceeded using two important
assumptions, and below we explain why these assumptions actually
follow from properties of two dimensional anyons.

\subsection{Properties of the particle exchanges}
Consider a two dimensional theory, in which particle of type $a$
is exchanged first with particle of type $b$ and then with
particle of type $c$. The operation should depend only on the type
of particle $a$, and the total topological charge of particles $b$
and $c$. The operation should not be able to distinguish the finer
splitting of the combined charge into the charges $b$ and $c$.
This condition can be summarized pictorially in Fig~\ref{fig:
forks}.

\begin{figure}[t]
\centerline{\includegraphics[width=1\columnwidth]{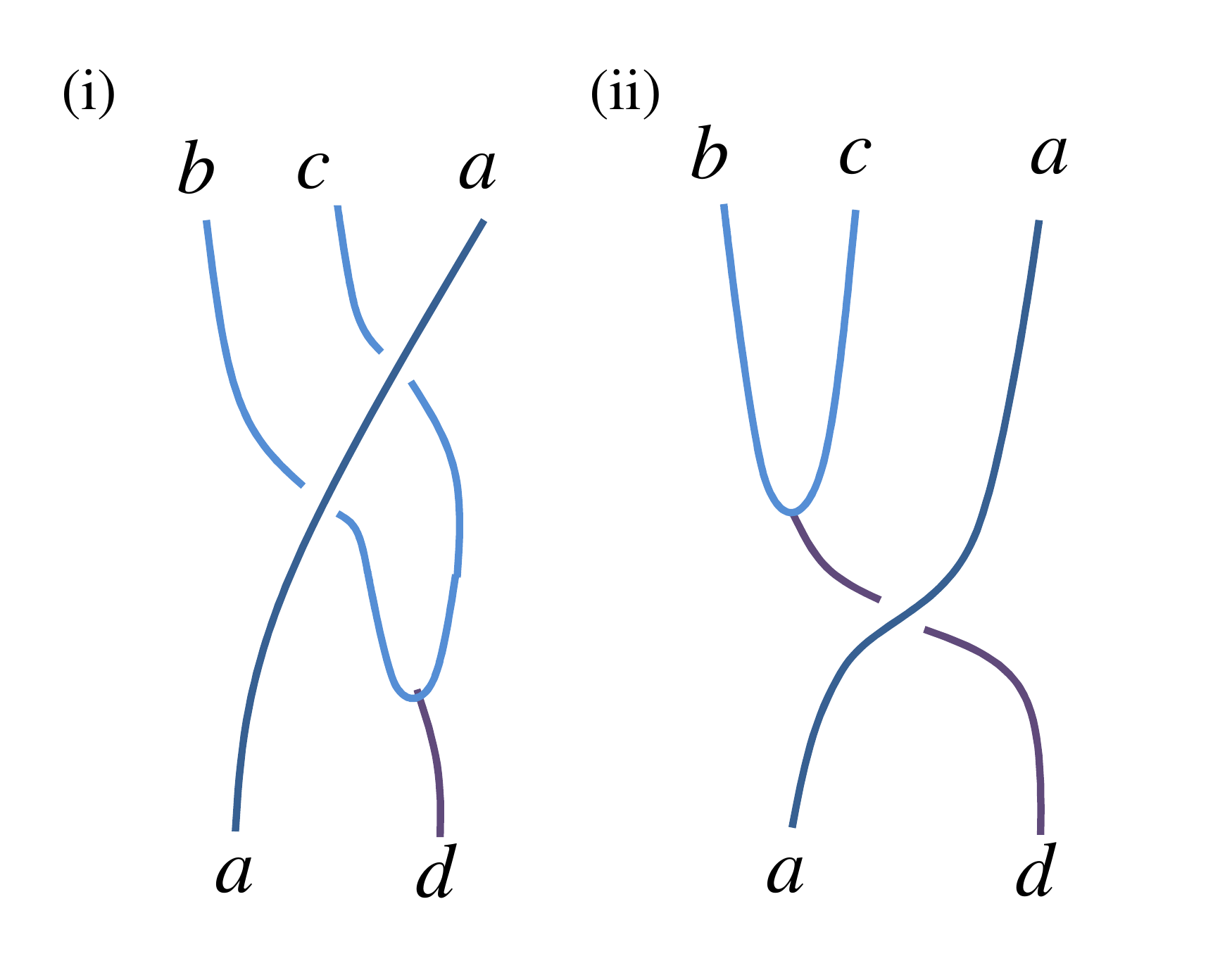}}
\caption{\textbf{Properties of particle exchanges} (i) Two consecutive  exchanges, first between particles $a$ and
$b$, and then between particles $b$ and $c$. (ii) Particle $a$ is
exchanged with a particle $d$, which is the particle resulting
from fusing $b$ and $c$. For two dimensional anyons, a proper
definition of the braid matrix between any two particle types
establishes an equality between (i) and (ii). For the case
$a=b=c=X$ and $d=q$, we only use the property that the two braids
in (i) move the label $q$ one segment to the left, as shown in
(ii). For the case where $a=q_1$,$b=q_2$,$c=q_3$ and $d=q_2+q_3
\mod 2m$, we show that the two sides are indeed equal}
\label{fig: forks}
\end{figure}

Importantly, the conditions summarized in Fig~\ref{fig: forks}
hold also in our one dimensional system, for the representation
derived in Sec.~\ref{sec: TS}, as we shall show below. The
importance of these conditions to our derivation in Sec.~\ref{sec:
TS} is threefold, since it can be used to (i) show that the
process in Fig.~\ref{fig: TS}(c) indeed exchanges the charges of
the two segments; (ii) show that the TS of the $q_1$, $q_2$
composite is equal to the TS of $q_1+q_2 \mod 2m$; (iii) show that
the Yang-Baxter equations hold for a braid of any two particle
types. Therefore, the only assumption necessary for the derivation
presented in Sec.~\ref{sec: TS} is that the condition summarized
in Fig.~\ref{fig: forks} holds.

In the following, we shall study three important cases, for which
we verify explicitly that the conditions in Fig.~\ref{fig: forks}
hold in our one dimensional system. These three cases are used to
verify the assumptions (i) and (ii) above.

\subsubsection{$a=b=c=X, d=q$}

For concreteness, Consider four neighboring interfaces
$X_1$--$X_4$, flanking alternating FM, SC and FM segments. The
operators corresponding to the charges in the different segments
are $\chi_{3\uparrow}\chi^{\dag}_{2\uparrow}=\q$,
$\chi_{2\uparrow}\chi^{\dag}_{1\uparrow}=\so$,
$\chi_{3\uparrow}\chi^{\dag}_{4\uparrow}=\stw$. We shall also use
$\chi_{4\uparrow}\chi^{\dag}_{1\uparrow}=\so\q\stw$. Consider an
initial state of the system which is an eigenstate of $\q$ with
eigenvalue $e^{i\frac{\pi}{m}q}$, and of $e^{i \pi \hat{S_1} +\hat
{S_2}}$ with eigenvalue $e^{i\frac{\pi}{m}s_{\rm tot}}$, \beq | q,
s_{\rm tot}\rangle =
\frac{1}{\sqrt{2m}}\sum_{n=0}^{2m-1}e^{i\frac{\pi}{m} q n}|s_1 =
-n, s_2=s_{\rm tot}+n\rangle \eeq where $s_1$ and $s_2$ correspond
to $\so$ and $\stw$.  Note that such a state is also an eigenstate
of $\chi_{4\uparrow}\chi^{\dag}_{1\uparrow}$.

Following Fig.~\ref{fig: forks}, consider two braid operations
first between $X_1$ and $X_2$ and then between $X_2$ and $X_3$.
 Using $\q|s_1,s_2\rangle=\vert
s_1-1,s_2+1\rangle$, and the Fourier representation of the braid
operators,  the resulting state is \bea
&\phantom{=}&(2m)\,e^{-i\frac{\pi}{2}}U_{12}(\hat{Q})U_{23}(\hat{S_1})| q, s_{\rm tot}\rangle \\
&=&\sum_{n,p,k=0}^{2m-1}e^{i\frac{\pi}{2m}\left[
-\left(p^2+k^2\right)+2 q n -2 n p\right]}| -n-k, s_{\rm
tot}+n+k\rangle \nonumber
\eea
Denoting $l=n+k$ we arrive at
\bea
&\phantom{=}&(2m)\,e^{-i\frac{\pi}{2}}U_{12}(\hat{Q})U_{23}(\hat{S_1})| q, s_{\rm tot}\rangle \\
&=&\sum_{n,p,l=0}^{2m-1}e^{i\frac{\pi}{2m}\left[
-\left((p+n)^2+l^2\right)+2 (q+l)n \right]}| -l, s_{\rm
tot}+l\rangle \nonumber \eea and therefore \bea
&\phantom{=}&(2m)\,e^{-i\frac{\pi}{2}}U_{12}(\hat{Q})U_{23}(\hat{S_1})| q, s_{\rm tot}\rangle \\
&=&\left(\sum_{r=0}^{2m-1}e^{-i\frac{\pi}{2m}r^2}\right)
e^{-i\frac{\pi}{2m}q^2}| s_1=q, s_2=s_{\rm tot}-q\rangle \nonumber
\label{eq: Xq} \eea We see therefore, that the two consecutive
exchanges are equivalent to moving the charge $q$ one segment to
the left, i.e., $e^{i\frac{\pi q}{m}}$ becomes the eigenvalue of
$\chi_{2\uparrow}\chi^{\dag}_{1\uparrow}=\so$.

That is exactly what is indicated in Fig.~\ref{fig: forks} (b).
The state is also multiplied by a phase which depends on the gauge
choices for the different basis \cite{KitaevHex}. Importantly,
note that we could have chosen to fuse $X_1$ with a different
interface $X_i$ (as long as it is not between $1$ and $3$), to
form a charge $\chi_{i\uparrow}\chi^{\dag}_{1\uparrow}$, which
would again commute with
$\chi_{3\uparrow}\chi^{\dag}_{2\uparrow}$. An identical analysis
to the above would yield the same result, with $\chi_{i\uparrow}$
replacing $\chi_{4\uparrow}$.

\subsubsection{Exchange of two $q$'s}
Using the above, we would now like to show that the four exchanges
depicted in Fig.~\ref{fig: TS} (c) indeed correspond to exchanging
two charges $q_1$ and $q_2$. Consider four neighboring interfaces
$X_0$--$X_3$, flanking alternating SC, FM, SC segments, where the
initial state is $\vert q_1, q_2\rangle$, corresponding to an
eigenstate of  $\chi_{1\uparrow}\chi^{\dag}_{0\uparrow}=\qo$ and
$\chi_{3\uparrow}\chi^{\dag}_{2\uparrow}=\qtw$.

Indeed, using the above results, we see that performing the
exchanges $X_1$ with $X_2$, and then $X_2$ and $X_3$, would result
in an eigenstate of $\chi_{2\uparrow}\chi^{\dag}_{1\uparrow}=\s$
with eigenvalue $e^{i\frac{\pi}{m}q_2}$. Now, performing the
exchanges $X_0$ with $X_1$, and then $X_1$ and $X_2$, and again
using the previous results, we see that the resulting state is an
eigenstate of $\chi_{1\uparrow}\chi^{\dag}_{0\uparrow}=\qo$ with
eigenvalue $e^{i\frac{\pi}{m}q_2}$. Therefore, it is an eigenstate
of $\chi_{3\uparrow}\chi^{\dag}_{2\uparrow}=\qtw$ with eigenvalue
$e^{i\frac{\pi}{m}q_1}$ (the total charge in the two segments is
preserved). Therefore, the two charges have been exchanged by the
sequence depicted in Fig.~\ref{fig: TS} (c).

We shall now explicitly calculate the phase factor resulting from
an exchange of two charges. Let us denote the operation at point
by $U_{\hat{Q}_1,\hat{Q}_2}$. We would now like to verify that
\bea
U_{\hat{Q}_1,\hat{Q}_2}\vert q_1,q_2\rangle &\equiv&
U_{23}(\hat{S})U_{34}(\hat{Q}_2)U_{12}(\hat{Q}_1)U_{23}(\hat{S})\vert
q_1,q_2\rangle\nonumber\\
&=&e^{i\phi(q_1,q_2)}\vert q_2,q_1\rangle,
\label{eq: verify}
\eea
and find the abelian phase $\phi(q_1,q_2)$ associated with this
exchange. The sequence of the four braid operation in the above
equation corresponds to Fig~\ref{fig: TS} (c). Using
$\s|q_1,q_2\rangle=\vert q_1-1,q_2+1\rangle$ and the Fourier
representation of the braid operators, we arrive at
\beq
\sum_{n,p=0}^{2m-1}e^{i\frac{\pi}{2m}\left[-\left(n^2+p^2\right)+\left(q_1-n\right)^2+\left(q_2+n\right)^2\right]}|q_1-n-p,q_2+n+p\rangle
\eeq
Denoting $l=p+n$ we arrive at
\beq
\sum_{n,l=0}^{2m-1}e^{i\frac{\pi}{2m}\left[\left(-l^2+2nl\right)+2\left(q_2-q_1\right)n+q_1^2+q_2^2\right]}|q_1-l,q_2+l\rangle
\eeq
The sum over $n$ forces $l=q_1-q_2$, which therefore gives
\beq
U_{\hat{Q}_1,\hat{Q}_2}\vert q_1,q_2\rangle=e^{i\frac{\pi}{m} q_1
q_2}|q_2, q_1\rangle.
\eeq
Therefore, Eq.~(\ref{eq: verify}) hold, and the abelian phase for
interchanging two charges $q_1$ and $q_2$ is just
$e^{i\frac{\pi}{m} q_1 q_2}$, which is consistent with the
topological spin for the charges, $\theta_q=e^{i\frac{\pi}{m}
q^2}$, Eq.~(\ref{eq: top spin}).

\subsubsection{$a=q_1$, $b=q_2$, $c=q_3$ $d=q_2+q_3 \mod 2m$}
Consider three consecutive SC segments, and an initial state
$\vert q_1,q_2,q_3\rangle$, corresponding to an eigenstate of
$\qo$,$\qtw$,and $\qt$. On the left side of Fig.~\ref{fig: forks},
we first have an exchange of the charge of the first and second
segments, resulting, according to the above discussion, in the
state $e^{i\frac{\pi}{m}q_1 q_2}\vert q_2,q_1,q_3\rangle$,
followed by an exchange of the second and third segment, resulting
in $e^{i\frac{\pi}{m}\left(q_1 q_2+ q_1q_3\right)}\vert
q_2,q_3,q_1\rangle$. Clearly, that is exactly the result of the
operation on the left side of Fig.~\ref{fig: forks}, in which the
total charge $q_2+q_3$ is exchanged with $q_1$ and then split
again into the two segments.

\subsection{Consistency check}
In this section we would like to verify that the topological spin
of the interfaces is indeed well defined, i.e., the phase acquired
by the operation defining the topological spin is independent of
the initial state of the system. Consider the procedure in
Fig.~\ref{fig: TS} (e). Interface $1$ flanks the left side of a SC
segment, which corresponds to $\q$. The initial state can be taken
as an eigenstate of $\q$ with eigenvalue $q$. A FM segment is
nucleated to the right of it. The FM segment is nucleated with
spin zero. Therefore, the state after the nucleation of this
segment is \beq
|\psi(q,s=0)\rangle=\frac{1}{\sqrt{2m}}\sum_{n=0}^{2m-1}|q_1=q-n,q_2=q+n\rangle
\eeq As explained in Sec.~\ref{sec: TS}, the braid operation
between interfaces $1$ and $2$ does not change the charge of the
SC segment between interfaces $1$ and $2$, the charge between $3$
and $4$, and therefore the total charge $q$ (numbering increases
to the right). Therefore, \beq
U_{12}|\psi(q,s)\rangle=\frac{1}{\sqrt{2m}}\sum_{n=0}^{2m-1}e^{i
\varphi(q-n)}|q_1=q-n,q_2=q+n\rangle \eeq where we keep the phase
function $\varphi(q)$ completely general. We now project on the
subspace with $\s=1$ in the FM segment, i.e. we apply the
projector \beq \Pi_{s=0}=\sum_q |\psi(q,s=0)\rangle
\langle\psi(s=0,q)| \eeq were the identity operation is implicitly
assumed to act on all other degrees of freedom. Applying the
projection yields (up to normalization), \beq
\theta_X=\sum_{n=0}^{2m-1} e^{i\varphi(q-n)}. \eeq Importantly,
$\theta_X$ does not depend on $q$, as the sum runs over all
possible values for charges.

\bibliography{fractional_refs}

\end{document}